\newcommand\bm[1]{\mbox{\boldmath $#1$}}
\newcommand\Eq[1]{Eq.$\,$(\ref{#1})}
\newcommand\Fig[1]{Fig.$\:$\ref{#1}}

\newcommand\w[1]{\mbox{$\omega_{#1}$}}

\newcommand\Id{\mbox{$\openone$}}

\newcommand\p{\mathrm{p}}

\newcommand{\squishlist}{
   \begin{list}{$\bullet$}
    { \setlength{\itemsep}{0pt}      \setlength{\parsep}{0pt}
      \setlength{\topsep}{5pt}       \setlength{\partopsep}{0pt}
      \setlength{\leftmargin}{1em} \setlength{\labelwidth}{1.em}
      \setlength{\labelsep}{0.3em} } }
\newcommand{\squishend}{
    \end{list}  }

\documentclass[aps,pra,reprint,groupedaddress,showpacs,amsmath,amssymb,floatfix]{revtex4-1}

\usepackage{graphicx}
\usepackage{bm}
\usepackage{amsmath}

\begin{document}



\title{A complete solution of the Bloch equation}

\author{Thomas E.~Skinner}
\email{thomas.skinner@wright.edu}  
\affiliation{Physics Department, Wright State University, Dayton, OH 45435}

\date{\today}

\begin{abstract}
The Bloch equation is the fundamental dynamical model applicable to arbitrary two-level systems.  Analytical solutions to date are incomplete for a number of reasons that motivate further investigation.  The solution obtained here for the propagator, which generates the time evolution of the system and embodies all the system dynamics, is compact and completely general.  The parameter space that results in division by zero in previous treatments is explicitly defined and accommodated in the solution.  Polynomial roots required for the solution are expressed in terms of a single real root obtained using simple functional forms.  A simple graphical rendition of this root is developed that clarifies and characterizes its dependence on the physical parameters of the problem. 
As a result, the explicit time dependence of the system as a function of its physical parameters is immediately evident.  Several intuitive models of system dynamics are also developed.  In particular, the Bloch equation is separable in the proper coordinate system, written as the sum of a relaxation operator and either a null operator or a commuting rotation.  The propagator thus drives either pure relaxation or relaxation followed by a rotation.  The paper provides a basis for increased physical insight into the Bloch equation and its widespread applications.

\end{abstract}



\maketitle

\section{Introduction}
The Bloch equation needs little formal introduction.  It was proposed originally as a classical, phenomenological model for the dissipative dynamics observed in magnetic resonance \cite{BlochEq}.  However, its impact has been more widespread as a result of its applicability to quantum two-level systems \cite{Feynman}. The field of quantum optics is a particularly noteworthy example of its significance beyond the realm of magnetic resonance \cite{Fano83}. More recently, the burgeoning field of quantum computing provides additional incentive for understanding Bloch equation dynamics due to the necessity of preserving coherence in the presence of relaxation \cite{ChuangQMC}.

A thorough treatment of this fundamental dynamical model for arbitrary 
two-level quantum systems, including relaxation, is therefore relevant to a host of important physical systems. Yet, there is to date no complete and general solution for the Bloch equation.  Previous solutions \cite{Torrey, KumarBlochEq1, KumarBlochEq2, BainBlochEq} share some or all of the following limitations.  They (i) divide by zero at particular, but unspecified, values of the field and relaxation rates; (ii) are not valid for arbitrary field orientation, which can be important in many applications; (iii) are cumbersome, linked to tables of multiply nested variables with obscure connection to the physical parameters of the problem; (iv) depend unavoidably on the roots of a cubic polynomial, roots that are only qualitatively specified or are expressed as complicated functions of the physical parameters; (v) provide only a small measure of the physical insight that might be expected from an analytical solution.

The present work derives a compact, complete solution to the Bloch equation. The solutions are valid for arbitrary (constant) fields and relaxation rates. The paper begins with a theoretical overview.
The intent is to provide a fairly complete general understanding of the problem and the formal simplicity of the solution.  The next section is devoted to the explicit form of the solutions. Their dependence on the physical parameters is fully characterized in a following section.  The spin-spin (transverse) relaxation rate minus the spin-lattice (longitudinal) rate provides a convenient and particularly useful frequency scale for representing system parameters in the analysis of the Bloch equation.  Conditions that result in division by zero in previous solutions are identified, yielding quantitative bounds for oscillatory (underdamped) and non-oscillatory (critically damped and underdamped) dynamics.  The polynomial roots required in the solution are reduced to a single root with a fully characterized dependence on the physical parameters that admits a simple graphical representation.   
Next, a simple vector model is developed that reveals the underlying simplicity of the dynamics, with a modified system of relaxation rates playing a role analogous to longitudinal and transverse relaxation. The modified rates result from the interaction/coupling between the fields and relaxation processes. Additionally, and incidentally, a method for finding eigenvectors emerges that does not appear to have been considered previously.  The Bloch equation is also shown to represent a system of three mutually coupled harmonic oscillators, providing additional perspective. Details of the calculations are provided in appendices.  The concluding appendix applies the solutions to a representative set of cases yielding solutions that can be straightforwardly verified by other methods.

\section{Theoretical Overview}

We first summarize the basic framework of the Bloch equation to recollect and also define the fundamental parameters of the problem.
The equation describes the dynamics of a magnetization $\bm{M}$ subjected to a static (polarizing) magnetic field $\bm{H}_0 = H_0\, \bm{\hat z}$ and a sinusoidally alternating field $2 H_a \cos\omega_a t$ applied orthogonal to $\bm{H}_0$. For $H_a \ll H_0$, the equilibrium magnetization is not appreciably affected by the applied field and is therefore, to a good approximation, the time-independent value $\bm{M}_0 = \chi H_0\,\bm{\hat z}$ produced by the polarizing field.  

One then considers a reference frame rotating about $\bm{H}_0$ at an angular frequency $\omega_a$ equal to the frequency of the applied field. In this frame, the resulting effective field $\bm{H}_e$ is also time-independent.  The evolution of the magnetization in this frame, neglecting dissipative effects, is simply a precession about the field at the Larmor frequency $\bm{\omega}_e = -\gamma\bm{H}_e$ due to the torque $\gamma\bm{M}\times\bm{H}_e$ on $\bm{M}$, with  
$\bm{H}_e = (H_a\cos\phi, H_a\sin\phi, H_0 - \omega_a/\gamma)$.  
The phase $\phi$ relative to the $x$-axis in the rotating frame is arbitrary in the context of a single applied field and can be set equal to zero.  However, the relative phase is required for problems involving sequentially applied fields. 

Relaxation rates $R_i$ are then assigned to each component $M_i$ to include dissipative processes.  The torque can be written as a matrix-vector product (Jaynes***), which, together with relaxation, gives the matrix 
     \begin{equation}
\Gamma =  \left( \begin{array}{ccc}
                       R_1 & \w{3} & -\w{2}  \\
                       -\w{3} & R_2 & \w{1}   \\
                       \w{2} & -\w{1} & R_3
                    \end{array}
             \right)
\label{Gamma}
     \end{equation}
comprised of the rates and the components of $\bm{\omega}_e$.  Including the initial polarization $M_0$ then gives the Bloch equation in the rotating frame as
     \begin{equation}
\bm{\dot M}(t) + \Gamma\bm{M}(t) = \bm{M}_{0}R_3.
\label{BlochEq}
     \end{equation}

In what follows, both $\bm{H}_e$ and $\bm{\omega}_e$ are referred to as fields, and we further define the transverse field $\bm{\omega}_{12}$ as a component of the total field $\bm{\omega}_e$, with respective magnitudes (squared)
     \begin{eqnarray}
\omega_{12}^2 &=& \omega_1^2 + \omega_2^2 \nonumber \\
\omega_{e}^2 &=& \omega_1^2 + \omega_2^2 + \omega_3^2.
\label{omega_mag}
     \end{eqnarray}
Framing the problem most generally to include arbitrary $\bm{\omega}_e$ and distinct relaxation rates might be expected to complicate the solution compared to previous treatments.  However, the symmetry introduced provides concise expressions for
$\Gamma_{ij} = \varepsilon_{ijk}\,\w{k}$ in terms of the usual Levi-Civita tensor (summed over repeated indices) and $\Gamma_{ii} = R_i$, resulting in a  compact and rather simple solution.  

\subsection{Bloch equation solution}

Multiplying \Eq{BlochEq} by $e^{\Gamma t}$ and integrating the resulting exact differential over the interval 
$[0,t\,]$ gives the solution
\begin{subequations}
  \label{BlochSol}
     \begin{eqnarray}
\bm{M}(t) &=&  e^{-\Gamma t}\bm{M}(0) + (1-e^{-\Gamma t})\bm{M}_\infty 
       \label{BlochSol:1} \\
          &=& e^{-\Gamma t}\,[\bm{M}(0) - \bm{M}_\infty] + \bm{M}_\infty
       \label{BlochSol:2}
     \end{eqnarray}
\end{subequations}
as a function of the steady-state 
$\bm{M}_\infty \equiv \Gamma^{-1}\bm{M}_0 R_3$ and transient
$\bm{M}(0)$ responses.  The crux of the problem is a solution for the propagator $e^{-\Gamma t}$.

\subsection{The propagator $\bm{e^{-\Gamma\,t}$}}
The Laplace transform $\mathcal{L}$ of $e^{-a t}$ is equal to ($s + a)^{-1}$ for constant $a$.  The matrix exponential 
$e^{-\Gamma t}$ for constant $\Gamma$ is then the inverse Laplace transform 
$\mathcal{L}^{-1}\,[\,(s\Id + \Gamma)^{-1}\,]$, where \Id\ is the identity element.  
The inverse Laplace transform of a function $f(s)$ can be written in terms of the Bromwich integral as (cf. \cite{Arfken})
    \begin{eqnarray}
\mathcal{L}^{-1}[f(s)] & = & \frac{1}{2\pi i}
     \int_{\gamma - i\infty}^{\gamma + i\infty} f(s) e^{s t}\,ds 
          \nonumber \\
 & = & F(t),
\label{Bromwich}      
     \end{eqnarray}
where the real constant $\gamma$ is chosen such that Re$\,(s) < \gamma$ for all singularities of $f(s)$.  Closing the contour by an infinite semicircle in the left half plane ensures convergence of the integral for $t>0$.  The desired
$F(t)$ is then the sum of the residues of the integrand.  

For $f(s) = (s\Id + \Gamma)^{-1}$, recall the textbook theorem for the inverse of a matrix $A$, with terms defined as follows:

\begin{itemize}
\item[(i)] $A(i|j)$ is the matrix obtained by deleting row $i$ and column $j$ of $A$.
\item[(ii)] The cofactor of $A_{ij}$ is $C_{ij} = (-1)^{i+j} \det A(i|j)$.
\item[(iii)] The adjugate of $A$ is the matrix
$(\mathrm{adj}\,A)_{ij} = C_{ji}$, i.e., the transpose of the cofactor matrix for $A$, which is the same as the cofactors of $A$ transpose.
\item[(iv)] Then $A^{-1} = \mathrm{adj}\,A/\det A$.
\end{itemize}

For
     \begin{equation}
A(s) = s\Id + \Gamma,
\label{adjA} 
     \end{equation}
the elements of 
$\mathrm{adj}\,A(s)$, are simple 
$(2 \times 2)$ determinants, giving 
     \begin{equation}
\mathrm{adj}\,A(s) = A_0 + A_1\, s + \Id \,s^2,
\label{adjA_Poly}
     \end{equation}
a polynomial in $s$ with coefficient matrices
     \begin{eqnarray}
(A_0)_{ij} & = & \w{i}\,\w{j}\, - \varepsilon_{ijk}\,\w{k} R_k
               + R_k R_m\, \delta_{ij} \nonumber \\
(A_1)_{ij} & = & (R_k + R_m)\,\delta_{ij} - \varepsilon_{ijk}\,\omega_k  \quad\quad\quad m\neq k\neq i \nonumber \\
\label{adjA_coeff}
     \end{eqnarray}
Thus, each element of $\mathrm{adj}\,A$ can be constructed from one of two relatively simple elementary forms, with diagonal or off-diagonal elements, respectively, related by cyclic permutation of indices. The matrices for the results above can be found in Appendix \ref{Vector Model}.  

The determinant of $A(s)$ is the characteristic polynomial of $(-\Gamma)$.  The resulting third degree polynomial $p\,(s)$ is of the form
     \begin{equation}
p\,(s) = c_{\,0} + c_1 s + c_2 s^2 + s^3 
\label{CharPoly}
     \end{equation}
with coefficients
     \begin{eqnarray}
c_{\,0} & = & \prod_i R_i + \sum_i R_i\, \omega_i^2 \nonumber \\
        & = & R_1 R_2 R_3 \left( 1 + {\displaystyle \sum_{i \ne j \ne k}} 
                         \dfrac{\omega_i^2}{R_j R_k}\,\right) \nonumber \\
c_1 & = & \omega_e^2 + R_1R_2 + R_1R_3 + R_2R_3 \nonumber \\
    & = & \omega_e^2 + \sum_{j<k} R_j R_k \nonumber \\
c_2 & = & \sum_i R_i\,.
\label{CPcoeff}
     \end{eqnarray}
One therefore has, simply,
     \begin{equation}
A^{-1}(s) = (s\Id + \Gamma)^{-1} = \dfrac{\mathrm{adj}\,A(s)}{p\,(s)}
\label{Ainv(s)}
     \end{equation}
              
The desired solution for
$F(t) = e^{-\Gamma t}$ is then the sum of the residues of the integrand in \Eq{Bromwich},
     \begin{equation}
e^{-\Gamma t} = \sum_\mathrm{res}\, 
                \dfrac{\mathrm{adj}\,A(s)}{p\,(s)}\,e^{s t}.
\label{F(t)}
     \end{equation}
Recall for reference in what follows that for a function $g(s)$ with a pole of order $k$ at $s=s_0$, the coefficient of $(s-s_0)^{-1}$ in the Laurent series expansion of $g(s)$ about $s=s_0$, i.e., the residue at $s_0$, is
     \begin{equation}
\mathrm{res\,}(s_0) = \dfrac{1}{(k-1)\,!}\lim_{s\rightarrow s_0}
                  \dfrac{d^{\,k-1}}{ds^{k-1}}\,[\,(s-s_0)^k\,g(s)\,]
\label{res}
     \end{equation}
The poles clearly occur at the roots of $p(s)$, i.e., the eigenvalues of $-\Gamma$.

\subsection{\textbf{Steady state solution}}
The steady state response $\bm{M}_\infty$ defined in \Eq{BlochSol} is equal to $\Gamma^{-1}\bm{M}_0 R_3$, with $\Gamma^{-1}$ obtained from \Eq{Ainv(s)} as $\textrm{adj}A(0)/p(0)$.  The dependence on $\textrm{adj}A$ is only in the third column, since $\bm{M}_0$ is along $\bm{\hat z}$, with $p(0)$ given by $c_0$ in \Eq{CPcoeff}.  Then
     \begin{eqnarray}
\bm{M}_\infty &=& 
   \frac{\chi H_0}
   {R_1 R_2\bigg(1 + {\displaystyle \sum_{i \ne j \ne k}} 
                         \dfrac{\omega_i^2}{R_j R_k}\,\bigg)}\,
   \left[\begin{array}{c}
         \omega_1\omega_3+\omega_2 R_2 \\
         \omega_2\omega_3-\omega_1 R_1 \\
         \omega_3^2 + R_1 R_2
         \end{array}
   \right]\,. \nonumber \\
\label{M_inf}
     \end{eqnarray}
Letting $R_1 = R_2 = 1/T_2$ and $R_3 = 1/T_1$ gives
     \begin{eqnarray}
 & & \nonumber \\
\bm{M}_\infty &\rightarrow& 
   \frac{\chi H_0}
   {1 + T_1 T_2\,\omega_{12}^2 + T_2^2\, \omega_3^2}
   \left[\begin{array}{c}
         T_2\,(\,\omega_1\omega_3 T_2 + \omega_2) \\
         T_2\,(\omega_2\omega_3 T_2-\omega_1) \\
         1 + T_2^2\, \omega_3^2
         \end{array}
   \right], \nonumber \\
\label{M_inf(T1,T2)}
     \end{eqnarray}
which reduces to Bloch's result \cite{BlochEq}, obtained for $\omega_2 = 0$.

\section{Solutions for the Propagator}
\label{PropSolns}

The results obtained so far provide the basis for a complete, compact, general solution of the Bloch equation.  The solution is valid for all values of the system parameters.  Degenerate roots of the characteristic polynomial, which give rise to division by zero in previous solutions, are fully addressed.  In a subsequent section, the solution provided in the most general form for the case of three unequal relaxation rates is applied to the more typical case $R_1 = R_2$ for isotropic media.  

\subsection{\textbf{Roots of the characteristic polynomial}}
\label{RootsCP}
The solution for $e^{-\Gamma\,t}$ given in \Eq{F(t)} requires the roots of the characteristic polynomial $p(s)$.  As is well known, the substitution 
$s = z -c_2/3$ reduces \Eq{CharPoly} to the standard canonical form
    \begin{eqnarray}
p\,(z-c_2/3) &=& z^3 + a z + b \nonumber \\
      &=& q\,(z), 
\label{SCForm}
     \end{eqnarray}
where
    \begin{eqnarray}
a & = & c_1 - c_2^2/3    \nonumber \\
b & = & 2\,\bigg(\dfrac{c_2}{3}\bigg)^3 -c_1\bigg(\dfrac{c_2}{3}\bigg) + c_0
\label{CardanCoeff}      
     \end{eqnarray}
Solutions for the roots $z_i$ are then available as functions of $a$ and $b$ from standard formulas.  However, the roots are complicated functions of the polynomial coefficients (and hence, the physical parameters in the Bloch equation), which hinders physical insight.  In Appendix \ref{AppPolyRoots}, simpler expressions are derived for the roots that reduce their complexity compared to previous treatments.
The fundamental results are summarized below.

Any polynomial with real coefficients has at least one real root, assigned here to $z_1$.  The solutions can then be consolidated in a convenient form that does not appear to have been employed before.  The other two roots are written as a function of $z_1$,  
     \begin{eqnarray}
z_{2,3} & \equiv & z_\pm \nonumber \\
       & = & -\frac{1}{2}z_1\, \pm \,i\,\varpi\, ,
\label{CubicRoots_z1}
     \end{eqnarray}
in terms of a discriminant 
     \begin{equation}
\varpi^2 = 3\big[(z_1/2)^2 + a/3\big],
\label{Discr}
     \end{equation}
which will be positive, negative, or zero depending on the value of $z_1$, the sign of $a$, and their relative magnitudes.  

The roots are further characterized here in terms of the positive parameter  
     \begin{equation}
\gamma  =  \dfrac{|b/2|}{|a/3|^{3/2}},
\label{gamma}
     \end{equation}
leading to the following delineation of the roots:
\squishlist
\item[(i)] $a>0$ or $a<0$ and $\gamma > 1$
   \squishlist
    \item[\qquad] 3 distinct roots (1 real, 2 complex conjugate)
   \squishend
\item[(ii)] $a<0$ and $\gamma<1$
   \squishlist
    \item[\qquad] 3 distinct real roots
   \squishend
\item[(iii)] $a<0$ and $\gamma=1$
   \squishlist
    \item[\qquad]  2-fold degenerate roots $z_+ = z_- = -\tfrac{1}{2}z_1$
   \squishend
\item[(iv)] $a = 0 = b$
   \squishlist
    \item[\qquad]  3-fold degenerate roots $z_i = 0$
   \squishend
\squishend
The physical parameters that define these effective domains for the roots are derived for the standard case $R_1 = R_2$ in a subsequent section.

In addition, the sign of $b$ determines the sign of $z_1$. Thus, in all cases, the set of three roots for a given $b<0$ is equal and opposite to the set obtained for parameters that flip the sign of $b$.  The case $b=0$ (i.e., $\gamma = 0$) reduces simply to $z_1 \sim \mbox{sgn}(0) = 0$.  From Eqs.~(\ref{CubicRoots_z1}) and (\ref{Discr}), there are then two additional real or imaginary roots depending on the sign of $\varpi^2$.  

The roots of $p\,(s)$ needed for the solution of $e^{-\Gamma\,t}$ in \Eq{F(t)} are then 
     \begin{equation}
s_i = z_i - c_2/3\, ,
\label{s_i}
     \end{equation}
where, referring to \Eq{CPcoeff},
     \begin{equation}
\frac{c_2}{3} = \frac{1}{3}\sum_i R_i \equiv \bar R
\label{Rbar}
     \end{equation}
is the average of the relaxation rates. 

\subsection{A convenient matrix partitioning}

We first seek to avoid transforming the characteristic polynomial to canonical form, solving for these roots, then transforming back to obtain the roots of the original polynomial. The result of this endeavor leads to additional simplifications in what follows.

Partition $\Gamma$ as the sum of commuting matrices
     \begin{eqnarray}
\Gamma & = & \mathcal{R} + \Gamma_\p  \nonumber \\
       & = & \bar R\, \Id + 
             \left( \begin{array}{ccc}
                       R_{1p}& \w{3} & -\w{2}  \\
                       -\w{3} & R_{2p} & \w{1}   \\
                       \w{2} & -\w{1} & R_{3p}
                    \end{array}
             \right),
\label{MatPartition}
     \end{eqnarray}
where the diagonal elements of $\Gamma_\p$ are
     \begin{eqnarray}
R_{ip} &=& R_i - \bar{R}  \nonumber \\
       &=& \frac{2}{3\,}R_i - \frac{1}{3}\,\sum_{j\ne i} R_j .
\label{R_ip}
     \end{eqnarray}
This partitioning gives $c_2 = \sum_i R_{ip} = 0$.  Therefore,
the characteristic polynomial for $-\Gamma_\p$ is in the standard canonical form $q(z)$ of \Eq{SCForm} 
with coefficients
    \begin{eqnarray}
a & = & \omega_e^2 + R_{1p}R_{2p} + R_{1p}R_{3p} + R_{2p}R_{3p}    \nonumber \\
b & = & \prod_i R_{ip} + \sum_i R_{ip}\, \omega_i^2
\label{CardanCoeff2}      
     \end{eqnarray}
from \Eq{CPcoeff}. 
We then have
     \begin{equation}
e^{-\Gamma t} = e^{-\bar R\, t} e^{-\Gamma_\p\, t}.
\label{MatExpP}
     \end{equation}

The focus henceforth will be the solution for $e^{-\Gamma_\p\, t}$, obtained from \Eq{F(t)} with roots $s_i=z_i$ given in \Eq{Root_z1} and 
$\mathrm{adj}\,A \rightarrow \mathrm{adj\,}A_\p $ obtained from $\Gamma_\p$.  Using Eqs.(\ref{adjA_Poly}) and (\ref{adjA_coeff}) with relaxation rates $R_i \rightarrow R_{i\p}$ gives
     \begin{eqnarray}
\mathrm{adj}\,A_\p(z) &=& A_{0\p} + A_{1\p}\, z + \Id \,z^2 \nonumber \\
(A_{0\p})_{ij} & = & \w{i}\,\w{j}\, - \varepsilon_{ijk}\,\w{k} R_{k\p}
               + R_{k\p} R_{m\p}\, \delta_{ij} \nonumber \\
(A_{1\p})_{ij} & = & (R_{k\p} + R_{m\p})\,\delta_{ij} - \varepsilon_{ijk}\,\omega_k  \quad\quad\quad m\neq k\neq i \nonumber \\
\label{adjAp_Poly}
     \end{eqnarray}

The partitioning also allows the coefficient matrices to be found rather easily in terms of $\Gamma_p$ as
     \begin{equation}
A_{0\mathrm{p}} = \Gamma_{\mathrm{p}}^2 + a\Id, \qquad A_{1\p} = -\Gamma_\p.
\label{adjAp_coeff_Gamma}
     \end{equation}
The result for $A_{1\p}$ is clear by inspection of the off-diagonal elements and confirmed using $\sum_i R_{ip} = 0$ for the diagonal elements.  One expands the $R_{i\p}$ according to \Eq{R_ip} to obtain the expression for $A_{0\p}$.  More generally, as can be verified by direct substitution,
     \begin{equation}
A_0 = c_1 - c_2\, \Gamma + \Gamma^2, \qquad A_1 = c_2 - \Gamma.
\label{adjA_coeff_Gamma}
     \end{equation}
The derivation is fairly straightforward and readily generalized to higher dimensional matrices.  However, these topics exceed the scope of the present work.

Either \Eq{adjAp_coeff_Gamma} or (\ref{adjA_coeff_Gamma}) can be used to obtain 
a concise implementation of the Cayley-Hamilton theorem, which states that every square matrix is a solution to its characteristic equation.  Thus, $\Gamma_\p$ is a solution of \Eq{SCForm}.  One can solve for $\Gamma_\p^3$, and subsequently for all higher powers of $\Gamma_\p$, in terms of the set $\{\Id, \Gamma_\p, \Gamma_\p^2\}$.  The series expansion of $e^{-\Gamma_\p\, t}$ can then be expressed in terms of the same set.  The net result is a relatively simple means for obtaining the scalar coefficients $a_i(t)$ in a solution of the form
     \begin{equation}
e^{-\Gamma t} = e^{-\bar R\, t}\, \big[\,a_0(t)\Id + a_1(t)\,\Gamma_\p + a_2(t)\,\Gamma_\p^2\,\big].
\label{MatExpC-H}
     \end{equation}

\subsection{Simple pole solution}
\label{Simple pole}
In the case that the roots $z_i$ of
$q\,(z)$ are distinct, the residues are due to simple first-order poles.  Factoring $q\,(z)$ as $\prod_i(z-z_i)$ gives 
$(z-z_n)/q\,(z)=\prod_{i \neq n}(z-z_i)$, as needed to evaluate the residue of a first order pole at $z=z_n$.  The derivative
$q^\prime(z) = \sum_j\prod_{i \neq j}(z-z_i)$ evaluated at $z_n$ is also
equal to $\prod_{i \neq n}(z_n-z_i)$, since the other terms in the sum vanish at $z = z_n$.  

The matrix exponential 
$e^{-\Gamma_\p t}$ can then be written simply and succintly as
     \begin{eqnarray}
e^{-\Gamma_\p t}\, &=& \sum_i 
       \frac{\mathrm{adj}\,A_{\p}(z_i)}{q^{\,\prime}(z_i)} e^{\,z_i t} 
\label{MatExp}
     \end{eqnarray}
There are thus independent contributions from each root, with shared dependencies on the fields $\omega_i$ and relaxation rates $R_{i p}$ which comprise $\mathrm{adj}\,A_\p$ according to \Eq{adjA_coeff}. 

Performing the sum, with substitutions from Eqs.\,(\ref{adjAp_Poly}) and (\ref{adjAp_coeff_Gamma}), and collecting terms according to \Eq{MatExpC-H} gives a result that can be written compactly as matrix multiplication in the form
     \begin{eqnarray}
e^{-\Gamma_\p t}\, &=& (\Id, \Gamma_\p, \Gamma_\p^2\,) 
    \left[
           \begin{array}{c}
               a_0(t) \\ a_1(t) \\ a_2(t)
           \end{array}  \right]  \nonumber \\
 & & \nonumber \\
        &=& (\Id, \Gamma_\p, \Gamma_\p^2\,)\, [\,W_1(z_1)\,\bm{u}_1(t)\,]\, , \nonumber \\ 
 &\ &          \nonumber \\  
W_1(z_1) &=& \left( 
    \begin{array}{ccc}
        z_1^2 + a \ \  & z_2^2 + 1 \ \  & z_3^2 + a \ \ \\
        -z_1 &- z_2\ \ & -z_3\ \  \\
        1\ \  & 1\ \  & 1
    \end{array}
             \right) \nonumber \\
 & &  \nonumber \\
\bm{u}_1(t) &=&  \left(
    \begin{array}{c}
       e^{z_1 t}/q^{\,\prime}(z_1) \\
       e^{z_2 t}/q^{\,\prime}(z_2) \\
       e^{z_2 t}/q^{\,\prime}(z_2)
    \end{array}
             \right)\, .
\label{MatExpSimplePole}
     \end{eqnarray}
The derivative of the characteristic polynomial can be calculated from either the factored form involving the roots or the polynomial form in \Eq{CharPoly}.  Each provides information that might be useful for different applications.

For parameter values
\squishlist
\item[(i)] $a>0$ or $a<0$ and $\gamma > 1$, 
\squishend
$\varpi$ is real from Eqs.~(\ref{Root_z1:1}) and (\ref{Root_z1:2}), so two of the roots are complex conjugates.
Although \Eq{MatExp} or (\ref{MatExpSimplePole}) is the most straightforward form of the solution and readily used in numerical calculations, the individual terms are complex.  A more transparently real-valued expression is obtained by performing the sum in \Eq{MatExp} after rationalizing complex denominators and writing the roots $z_{2,3}$ in terms of $z_1$ using Eqs.\,(\ref{CubicRoots_z1}) and (\ref{Discr}), as detailed in Appendix \ref{App:CalcProp}.  The result is \Eq{MatExpSimplePole} with
     \begin{eqnarray}   
W_1(z_1) &\rightarrow&  \frac{1}{3z_1^2 +a}\, \left( 
    \begin{array}{ccc}
        z_1^2 + a \ \  & 2z_1^2 \ \  & -a z_1  \\
        -z_1 & z_1 \ \ & -(\tfrac{3}{2}z_1^2 + a)  \ \  \\
        1\ \  & -1\ \  & -\tfrac{3}{2}z_1
    \end{array}
             \right) \nonumber \\
 & &  \nonumber \\
\bm{u}_1(t) &\rightarrow&  \left( 
    \begin{array}{c}
       e^{z_1 t} \\
       e^{-z_1 t/2}\, \cos\varpi t \\
       e^{-z_1 t/2}\, \dfrac{\sin\varpi t}{\varpi}
    \end{array}
             \right)\, .
\label{MatExp1}
     \end{eqnarray}
The coefficient $a$ can be found in terms of the roots $z_i$ upon expanding the factored form for $q(z)$ to obtain $a = z_1 z_2 + z_1 z_3 + z_2 z_3$.
The solution for the matrix exponential is thus separable into a term the depends directly on the physical parameters of the problem through $\Gamma_p$, a term that depends on the roots $z_i$, and a term that gives the time dependence, which in turn is solely a function of the roots.

For the case
\squishlist
\item[(ii)] $a<0$ and $\gamma < 1$, 
\squishend
$\varpi$ is imaginary, as given by \Eq{Root_z1:3}, so there are three real roots.  There is no oscillatory behavior in the straightforward result given in \Eq{MatExpSimplePole}.  
The solution can written alternatively in terms of $\mu = |\varpi|$ using \Eq{MatExp1}, with $\varpi = i\mu$ giving 
$\cos\varpi\,t \rightarrow \cosh\mu\,t$ and
$\sin\varpi\,t/\varpi \rightarrow \sinh\mu\,t/\mu$.

\subsection{Second-order pole solution}
For the case
\squishlist
\item[(iii)] $a<0$ and $\gamma = 1$, 
\squishend
two of the three real roots are equal, giving a doubly degenerate root 
$z_2 = z_3 = -z_1/2$, since $\varpi = 0$ when $\gamma = 1$ in either \Eq{Root_z1:2} or
\Eq{Root_z1:3}.   The characteristic polynomial
$q\,(z)\rightarrow (z-z_1)(z-z_2)^2$.  The contribution from the first-order pole at 
$z_1$ is obtained as before from the $i=1$ term of \Eq{MatExp}. The residue at $z_2$ is calculated in Appendix \ref{App:CalcProp}, leading to a solution that can be written in the form
     \begin{eqnarray}
e^{-\Gamma_\p t}\, &=& (\Id, \Gamma_\p, \,\Gamma^2_\p\,)\, 
                       [\,W_2(z_2)\,\bm{u}_2(t)\,]\, , \nonumber \\ 
 & &          \nonumber \\       
W_2(z_1) &=&  \left( 
    \begin{array}{ccc}
       \dfrac{1}{9}   & \dfrac{8}{9}  & \dfrac{1}{3}z_1  \\
         & & \\
        -\dfrac{4}{9}z_1^{-1} & \dfrac{4}{9}z_1^{-1}  & -\dfrac{1}{3}  \\
         & & \\  
        \dfrac{4}{9}z_1^{-2} & -\dfrac{4}{9}z_1^{-2} & 
                         -\dfrac{2}{3}z_1^{-1}
    \end{array}
             \right) \nonumber \\
 & &  \nonumber \\
 & &  \nonumber \\
\bm{u}_2(t) &=&  \left( 
    \begin{array}{c}
       e^{z_1 t} \\
       e^{-z_1 t/2} \\
       t e^{-z_1 t/2}
    \end{array}
             \right)\, .
\label{MatExp2}
     \end{eqnarray}
There is thus a term linear in the time, $t$.
Note that \Eq{MatExp2} is also the limit of \Eq{MatExp1} as 
$\varpi \rightarrow 0$, which requires $a \rightarrow -3(z_1/2)^2$ according to \Eq{Discr}, providing an independent verification of the simple-pole result.

\subsection{Third-order pole solution}

The case
\squishlist
\item[(iv)] $a=0=b$ 
\squishend
gives a triply degenerate, real root $z_1 = 0$ for 
$q\,(z) \rightarrow z^3$. The residue 
is one-half the second derivative of $\mathrm{adj}\,A_\p(z)\,e^{z t}$ with respect to $z$, evaluated at $z=0$, giving
    \begin{eqnarray}
e^{-\Gamma_\p t}\, &=&  \frac{1}{2}\,(\mathrm{adj}A_\p)^{\,\prime\prime}\big|_{0}
           + t\,(\mathrm{adj}A_\p)^{\,\prime}\big|_{0} + 
             \frac{t^2}{2}\,\mathrm{adj}A_\p(0)  \nonumber \\
  &=& \Id + A_{1\p}\,t + \tfrac{1}{2}A_{0\p}\, t^2 \nonumber \\
  &=&  \Id - \Gamma_\p\,t + \tfrac{1}{2}\Gamma_\p^2\, t^2\,.
\label{MatExp3}      
     \end{eqnarray}
There is now a term that is quadratic in the time. The same result is obtained from \Eq{MatExp2} in the limit 
$z_1 \rightarrow 0$ upon series expansion of the exponential terms.   

In addition, the Cayley-Hamilton theorem is simple to apply directly in this case, since $q(\Gamma_\p) = 0 = \Gamma_\p^3$.  The series expansion of $e^{-\Gamma_\p\,t}$ is therefore truncated, giving the \Eq{MatExp3} result directly and verifying the self-consistency of the solutions.

As mentioned at the beginning of the section, the solutions can be further simplified when $R_1 = R_2$ to provide increased insight into the nature of the solutions and the constraints that determine root multiplicities.

\section{Characterization of the Solutions}
\label{R1=R2 Solutions}

Substituting $R_1 = R_2$ gives rates $R_{ip}$ which can be written in the simple form
     \begin{equation}
R_{1p} = R_{2p} = R_\delta, \qquad R_{3p} = -2R_\delta, 
\label{R_ip:R1=R2}
     \end{equation}
where
     \begin{equation}
R_\delta = \frac{R_2-R_3}{3} \geq 0, 
\label{R_del}
     \end{equation}
since the transverse relaxation rate $R_2$ is greater than or equal to the longitudinal rate $R_3$ in physical systems.
The coefficients of the characteristic polynomial for $-\Gamma_\p$ then simplify to
    \begin{eqnarray}
a & = & \omega_e^2 - 3R_\delta^2    \nonumber \\
b & = & R_\delta
       \big[\,\omega_e^2 - 2\,R_\delta^2- 3\,\omega_3^2\,\big]\,.
\label{CardanCoeff3}      
     \end{eqnarray}
The rate $R_\delta$ provides a convenient and simplifying frequency scale for characterizing the solutions in the sections which follow.
  
\subsection{Criteria for the existence of degenerate roots}
\label{DegRoots}
The resulting simpler form for the polynomial coefficients makes possible a straightforward analysis of the conditions for which there are degeneracies in the roots.  As discussed in section \ref{RootsCP}, there is a two-fold degeneracy in the roots for $a<0$ and $\gamma=1$.  This is equivalent, using \Eq{gamma} for $\gamma$, to 
     \begin{eqnarray}
D(a,b) &=& (b/2)^2 + (a/3)^3  \nonumber \\
       &=& 0.
\label{Dfunc}
     \end{eqnarray}
The trivial solution $a=0=b$ gives a three-fold degenerate root $z_i=0$.

Details are deferred to Appendix \ref{AppDegRoots}, where the existence of degenerate roots is characterized in terms of
    \begin{equation}
\omega_3^2 = \lambda_3 R_\delta^2/3  \qquad \mathrm{and}  \qquad
             \omega_{12}^2 = \lambda_{12} R_\delta^2/3 .
\label{Param_w}      
     \end{equation}
For each $\omega_3$ defined by the range $0 \le \lambda_3 \le 1$, one finds two solutions for $\lambda_{12}$ that satisfy $D(a,b) = 0$ and give real values for $\omega_{12}$. Thus, for each $\omega_3 \in [\,0,R_\delta^2/3\,]$, there are two values of $\omega_{12}$  that produce degeneracies in the roots $z_i$.  The two solutions for $\lambda_{12}$ can be expressed concisely in the form 
    \begin{eqnarray}
\lambda_{12,i} &=& \eta_i - \lambda_3 + \tfrac{9}{4}  
    \qquad\qquad\qquad
            i=1,2  \nonumber \\
\eta_i &=& \tfrac{9}{2}\sqrt{8\lambda_3 + 1} \, \sin\vartheta_i 
       \nonumber \\
     & &  \nonumber \\
\vartheta_1 &=& \mathrm{sgn}(\lambda_3 - \lambda_b)\, \tfrac{1}{3}\sin^{-1}\frac{|8\lambda_3^2 + 20\lambda_3 -1|}{(8\lambda_3 + 1)^{3/2}} \nonumber \\
\vartheta_2 &=& \pi/3 - \vartheta_1 
\label{etaSolns}      
     \end{eqnarray}
for $\lambda_b =  \tfrac{3}{4}(\sqrt{3} - \tfrac{5}{3})$.  The solutions converge at $\lambda_3 = 1$ to $\eta_1 = \eta_2 = 27/4$, 
giving $\omega_{12}^2 = 8(R_\delta^2/3)$.  Then $a = 0 = b$ from \Eq{CardanCoeff2}, giving a three-fold degenerate root $z_i=0$ of \Eq{SCForm}.

The following simple and explicit criteria define the characteristics of the roots:
\squishlist
\item[(i)] $\omega_3^2 > R_\delta^2/3$
\squishend
there is no real-valued solution for $\omega_{12}^2$ 
such that $\gamma = 1$, i.e., $D(a,b)=0$,
and hence no degenerate roots $z_i$.  One then has the simple-pole solution of \Eq{MatExp1}.
\squishlist
\item[(ii)] $\omega_3^2 < R_\delta^2/3$
\squishend
there are two different real-valued solutions for $\omega_{12}^2$ as a function of $\lambda_3 $
that each give a two-fold degeneracy in the roots $z_i$, requiring the second-order pole solution of \Eq{MatExp2}.
\squishlist
\item[(iii)] $\omega_3^2 = R_\delta^2/3$
\squishend
gives $\omega_{12}^2 = 8(R_\delta^2/3)$ for $\lambda_3 = 1$, resulting in a three-fold degenerate root $z_i=0$ which requires the third-order pole solution of \Eq{MatExp3}.

\subsection{\textbf{Characterization of the damping}}
Solutions for the roots $z_i$ are characterized according to whether the discriminant $\varpi^2$ of \Eq{Discr} is positive, negative, or zero, and can be described, respectively, as underdamped, overdamped, or critically damped, analogous to a damped harmonic oscillator.  

\begin{figure*}[h]
\includegraphics[scale=1]{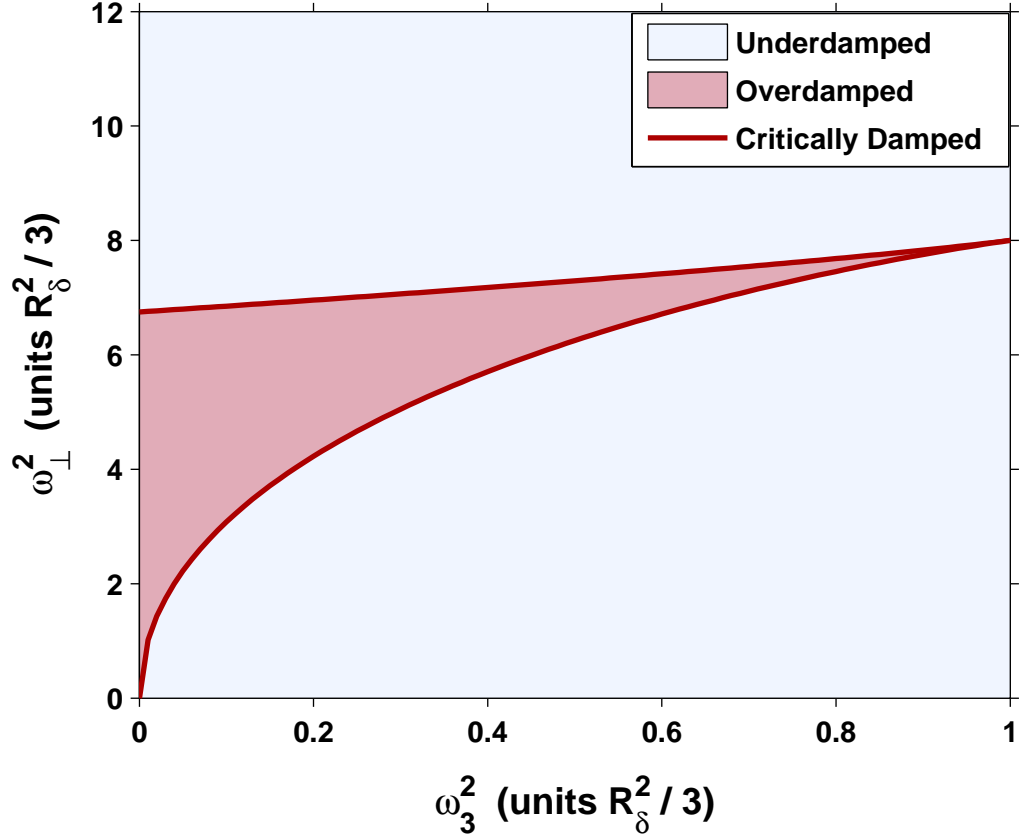}
\caption{Parameter values of $\omega_{12}^2$ that give degenerate roots of the characteristic polynomial ($a < 0, \gamma = 1$) and critically damped solutions to the Bloch equation are plotted as a function of $\omega_3^2$, shown as red (solid) lines calculated using \Eq{etaSolns}.  The parameters are scaled to $R_\delta^2/3$ as in \Eq{Param_w}.  In the interior of the region delineated by these curves (light red), there are three distinct real roots ($a < 0, \gamma < 1$) resulting in overdamped solutions.  Outside this region (light blue), one real and two complex conjugate roots  produce oscillatory, underdamped solutions, with $a > 0$ above the overdamped region and $a > 0, \gamma > 1$ below the overdamped region.  
}
\label{RootsDomain}
\end{figure*}

The solution for the propagator in the case of degenerate roots ($a<0, \gamma=1$) has a term linear in time, characteristic of a critically damped harmonic oscillator.  For a three-fold degeneracy in the roots, there is an additional term that is quadratic in the time.  The allowed values of $\omega_3^2$, as discussed in the previous section, are restricted to the narrow range parameterized according to $0\le\lambda_3 \le 1$.  The two solutions $\omega_{12,1}^2$ and $\omega_{12,2}^2$ for each $\omega_3^2$, as determined from Eqs.~(\ref{Param_w}) and (\ref{etaSolns}), are the solid curves plotted in \Fig{RootsDomain}.  

Using the same scaling of $\omega_3$ and $\omega_{12}$ as in \Eq{Param_w}, we also have
    \begin{eqnarray}
a(\lambda_{12},\lambda_3) &=& (\lambda_{12} + \lambda_3 - 9)\,R_\delta^2/3 \nonumber \\
b(\lambda_{12},\lambda_3) &=& (\lambda_{12} - 2\lambda_3 - 6)\,R_\delta^3/3 \nonumber \\
\gamma(\lambda_{12},\lambda_3) &=& \frac{9}{2}\, 
        \frac{|\lambda_{12} - 2\lambda_3 - 6|}{|\lambda_{12} + \lambda_3 - 9|^{3/2}}
\label{a,b,g scaling}      
     \end{eqnarray}
Solutions in the range $\omega_{12,1}^2 < \omega_{12}^2 < \omega_{12,2}^2$ bounded by the critical damping parameters give
$a<0$ and $\gamma<1$, resulting in three distinct real roots and overdamped evolution.  The range of bounding values is fairly narrow, becoming increasingly so with increasing $\lambda_3 $ and converging to a single value 
$\omega_{12}^2 = 8 R_\delta^2/3$ as $\lambda_3 \rightarrow 1$, as shown in the figure.

Underdamped, oscillatory solutions are obtained for all other field values, either
$\omega_3^2 > R_\delta^2/3$ (i.e., $\lambda_3 > 1$) or 
$\omega_{12}^2 \ge \omega_{12,1}^2$ and $\omega_{12}^2 \le \omega_{12,2}^2$ 
for $\lambda_3 \le 1$.

\subsection{\textbf{Characterization of the roots}}
The solution to the Bloch equation has a relatively simple form and can be expressed in terms of a single root, $z_1$, of the characteristic polynomial for $-\Gamma_p$.  Although the solutions for $z_1$ have also been expressed in relatively simple functional form, these forms provide little physical insight.  It remains to shed some light on the dependence of this root on the field $\bm{\omega}_e$ and the relaxation rates. 

\subsubsection{\bf{Physical limits of the roots}}

The roots $z_i$, being functions of $a, b$ and $\gamma$, also scale as $R_\delta$.  The associated decay rates are 
$\mathrm{Re}(s_i) = \mathrm{Re}(z_i) - \bar R$, from \Eq{s_i}.  Defining
     \begin{equation}
\lambda_z = \mathrm{Re}(z_i)/R_\delta.
\label{Lambda_z}
     \end{equation}
and using \Eq{R_del} for $R_\delta$ gives the decay rates
     \begin{eqnarray}
\mathrm{Re}(s_i) &=& \lambda_z R_\delta - \bar R \nonumber \\
                 &=& -\frac{(2-\lambda_z)}{3} R_2 
                                 -\frac{(1+\lambda_z)}{3} R_3  .
\label{DecayRates}
     \end{eqnarray}
The limiting rates are $R_2$ and $R_3$, which therefore constrains 
$\lambda_z$ to the range
     \begin{equation}
-1 \le \lambda_z \le 2.
\label{z1 Range}
     \end{equation}
The damping has equal contributions from $R_2$ and $R_3$ for $\lambda_z = 1/2$, with a larger contribution from either $R_2$ or $R_3$ if $\lambda_z$ is less than or greater than 1/2, respectively.

\begin{figure*}[h]
\includegraphics[scale=.85]{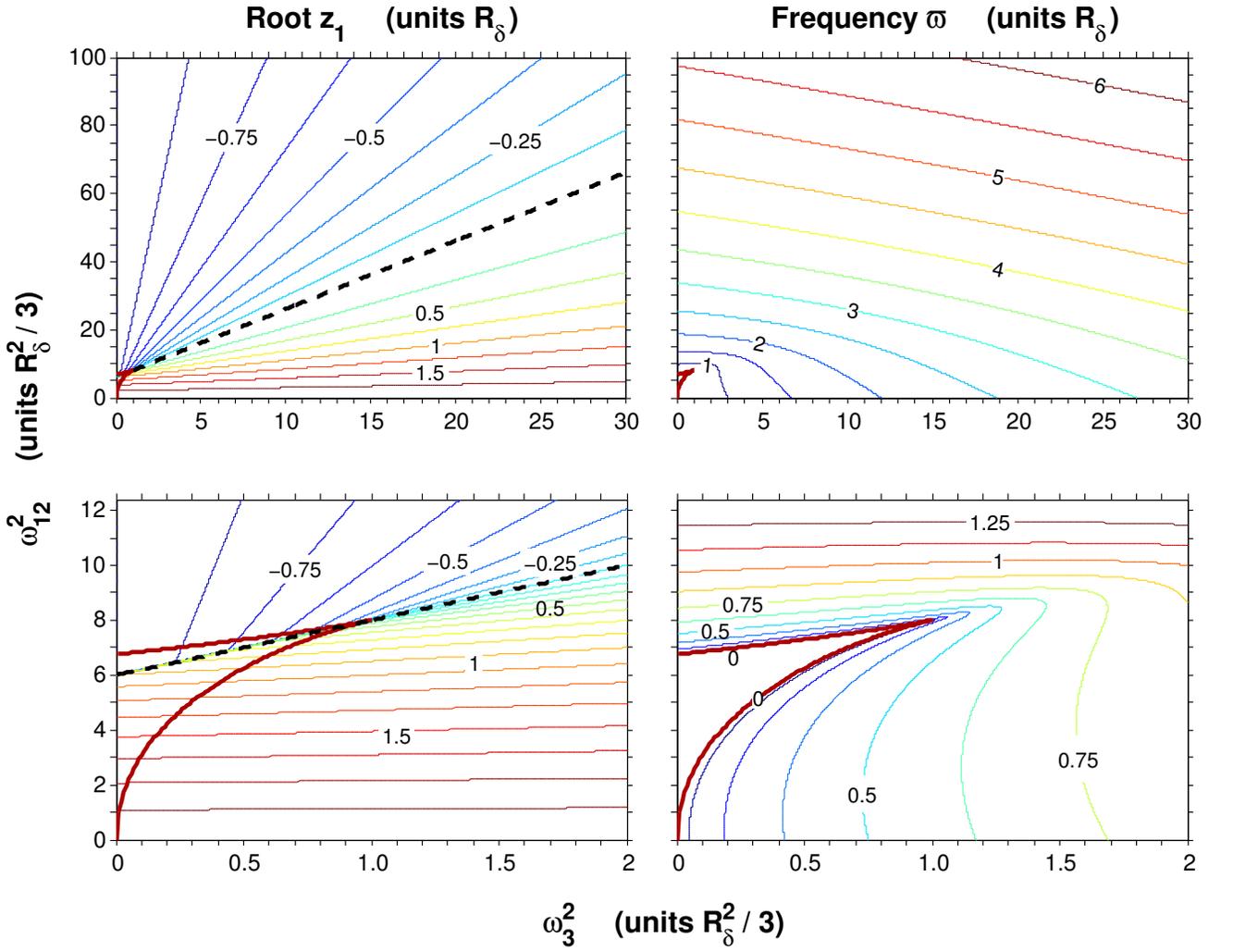}
\caption{Contours of the characteristic polynomial's guaranteed real root $z_1$, calculated according to Eqs.\,(\ref{Root_z1}) and normalized to $R_\delta$, are plotted as a function of $\omega_{12}^2$ and $\omega_3^2$ normalized as in \Fig{RootsDomain}.  The root satisfies
$-1 \le z_1 \le 2$, as expected from \Eq{DecayRates}, with lines of constant $z_1$ as derived in Eqs.\,(\ref{Linear b(a)}--\ref{Slope, Int}).  The $z_1 = 0$ contour is shown as a dashed line. Contours of the frequency $\varpi$ from \Eq{Discr} that appears in the oscillatory, underdamped solutions of the Bloch equation are also plotted in the rightmost panels.  Within the overdamped region defined in \Fig{RootsDomain} and expanded in the lower panels, there is no oscillation or frequency $\varpi$, and only one of the three real roots is plotted.
}
\label{z1_w}
\end{figure*}

The dependence of $z_1$ on $\bm{\omega}_e$ and $R_\delta$, calculated according to Eqs.$\,$(\ref{Root_z1}), is shown in 
\Fig{z1_w}, where contours of $\lambda_{z}$ are plotted as a function of $\lambda_{12}$ and $\lambda_3$. As discussed earlier, there is only one real root for $\lambda_3 > 1$.  When $\lambda_3 \le 1$, there is also a single real root for values of $\lambda_{12}$ outside the narrow bounds that define critical damping.  Within these bounds where the solutions represent overdamping, any of the three real roots can be designated as $z_1$, with $z_\pm$ from \Eq{Root_z1:3} giving the other two. 
For $\omega_{12} = 0$, the relaxation rate is $R_3$ (i.e., $\lambda_{z} = 2$), independent of the offset parameter $\lambda_3$, as is well-known.  As $\omega_{12}$ increases for fixed $\omega_3$, the relaxation rate approaches $R_2$ ($\lambda_{z} = -1$), with the drop-off from $\lambda_z = 2$ becoming increasingly steep at lower values of $\omega_3$.  For the other roots in which $\mathrm{Re}(z_\pm) = -1/2\, z_1$, the upper limit in 
\Eq{z1 Range} becomes 1/2. 

\subsubsection{\bf{A linear relation for the roots}}

Equation (\ref{SCForm}) evaluated at the real root $z_1$ yields the linear relation
     \begin{equation}
b = -z_1 a - z_1^3
\label{Linear b(a)}
     \end{equation}
for coefficient $b$ that will satisfy \Eq{SCForm} as a function of a given coefficient $a$, with slope and intercept determined by $z_1$.  Substituting the expressions for $a$ and $b$ in \Eq{a,b,g scaling}, rearranging and collecting terms after writing $9\lambda_z = 6\lambda_z + 3\lambda_z$ gives
     \begin{equation}
\lambda_{12} = m_\mathrm{s}\, \lambda_3 + \lambda_{12}^{\mathrm{int}}
\label{Linear Lam_12(Lam_3)}
     \end{equation}
with slope $m_\mathrm{s}$ and intercept $\lambda_{12}^{\mathrm{int}}$
     \begin{equation}
m_\mathrm{s} = \frac{2-\lambda_z}{1+\lambda_z}, \qquad\quad
               y_{12}^{\mathrm{int}} = 3(2-\lambda_z)(1+\lambda_z).
\label{Slope, Int}
     \end{equation}
There is thus a simple graphical representation for the value of the root $z_1$ as a function of the physical parameters $\,\omega_{12}, \omega_3, R_\delta$.  There are a continuum of field values for a given $R_\delta$ that give the same $z_1$.  Lines of constant $z_1$ as a function of $\lambda_{12}$ and $\lambda_3$ become hyperbolas when 
\Eq{Linear Lam_12(Lam_3)} is rewritten in terms of $\omega_{12}^2, \omega_3^2, R_\delta^2$ using \Eq{Param_w}. 

\section{\textbf{Intuitive Representations of System Dynamics}}
\label{Propagator Dynamics}
In most cases, the parameters of the Bloch equation yield three distinct roots for the characteristic polynomial $p(s)$ of \Eq{CharPoly}, described as cases (i) and (ii) in section \ref{RootsCP}.  Exceptions were considered in more detail in section \ref{R1=R2 Solutions} for the condition $R_1 = R_2$.  To provide additional physical insight, we develop a straightforward vector model of the time evolution for $M(t)$ given in \Eq{BlochSol}.  This requires the action of the propagator $e^{-\Gamma\,t}$ on an arbitrary vector.  An alternative vector model is also considered, followed by a coupled oscillator model.

The eigensystem for $\Gamma$ is considered in sections that follow, but one can substitute notation for the partitioned matrix $\Gamma_\p$ in the expressions which are derived, since, as defined in \Eq{MatPartition}, the matrices differ by a constant $\bar R$ times the identity matrix.  The difference in the eigenvalues is also $\bar R$, from Eqs.\,(\ref{s_i}) and 
(\ref{Rbar}).  Thus $-\Gamma$ and $-\Gamma_\p$ have the same eigenvectors $\bm{s}_i \equiv \bm{z}_i$. Simple analytical expressions for the eigenvectors and other constituents of the model are derived in Appendix \ref{Vector Model}. Each (unnormalized) eigenvector, which can assume different analytical forms depending on the scaling, is found to comprise the columns of $\mathrm{adj}\,A(s_i) = \mathrm{adj}\,A_\p(z_i)$, as discussed in Appendix \ref{EigVecCalc}, providing an alternative method for calculating an eigenvector.  

\subsection{Existing models specific to simple limiting cases}

As a point of departure, consider first the simple limiting cases for which the dynamics is already well known and readily visualized.
In the absence of relaxation, i.e., all $R_i = 0$, any magnetization vector $\mathcal{M}$ rotates about the total effective field $\bm{\omega}_e$ at constant angular frequency $\omega_e$. The time evolution of a vector under the action of the propagator has a simple solution in a coordinate system rotated to align one of the axes with the effective field.  The component of $\mathcal{M}$ along $\bm{\omega}_e$ is constant, and the components in the plane perpendicular to $\bm{\omega}_e$ rotate at angular frequency $\omega_e$ in the plane.  By constrast, the solution for each component $\mathcal{M}_i(t)$ in the standard $(x_1,x_2,x_3)$-coordinate system is more complicated, and it is not immediately apparent by inspection that the solution is a rotation.  

If the relaxation is switched on with equal rates $R_i = R$ on the diagonal, the relaxation matrix $R \Id$ commutes with the remaining rotation matrix, and the solution is a dynamic scaling $e^{-R t}$ of the rotating vector $\mathcal{M}$.  In addition, for $\omega_{12}=0$ and $R_1 = R_2 \neq R_3$, the relaxation matrix still commutes with the rotation about nonzero $\omega_3$.  The evolution is then a scaling $e^{-R_2 t}$ of the transverse component $\mathcal{M}_{12}$, which rotates at angular frequency $\omega_3$ in the plane perpendicular to $\omega_3$, along with exponential decay $ e^{-R_3 t}$ of component $\mathcal{M}_3$, as illustrated in \Fig{VMFigs}a.  In the case of pure relaxation, with all the field components 
$\omega_i = 0$, the solution is a non-oscillatory exponential decay 
$e^{-R_i t}$ for each component $\mathcal{M}_i$ along coordinate axis $x_i$. 

\subsection{A more general model}
\label{VM}
With the exception of the above simple cases, there has been no analogous picture of system dynamics when the rotation and relaxation do not commute.  The combined, noncommutative action of arbitrary fields and dissipation rates appears to require something more complex. Yet, the simple visual model shown in \Fig{VMFigs}a, which is comprised of independent relaxation and rotation elements, is readily extended to the general case when viewed in an appropriate coordinate system.  

\subsubsection{\bf{One real, two complex conjugate roots}}
\label{1R2C roots}

The solution for each component $\mathcal{M}_i$ is known to be a combination of oscillation and bi-exponential decay \cite{Torrey}, as is also evident from  the propagator derived in \Eq{F(t)}.  The underlying simplicity of the system dynamics can be demonstrated starting with the eigensystem for $\Gamma$ (or, alternatively, $\Gamma_\p$, as noted above). 

The real eigenvalue $s_1$ of $-\Gamma$ has a real eigenvector $\bm{s}_1$ which can be used as one axis of a physical coordinate system, but the complex roots $s_+$ and $s_- = s_+^*$ have associated complex eigenvectors $\bm{s}_+$ and $\bm{s}_- = \bm{s}_+^*$.  The eigenvectors are most generally not orthogonal, but they are linearly independent, given the distinct eigenvalues.  

Define the real vectors
     \begin{eqnarray}
\bm{\tilde{s}}_1 = \bm{s}_1, \qquad
      \bm{\tilde{s}}_2 &=&  \tfrac{1}{2}\,(\bm{s}_+ + \bm{s}_-)\, ,\qquad 
\bm{\tilde{s}}_3 = -\frac{i}{2}\,(\bm{s}_+ - \bm{s}_-)  \nonumber \\
                       &=& \mathrm{Re}\,[\bm{s}_+]\,, \hskip .75truein
                 = \mathrm{Im}\,[\bm{s}_+]\,.  \nonumber \\
& &
\label{Real s_2,3}
     \end{eqnarray}
The set $\{\bm{\tilde{s}}_1, \bm{\tilde{s}}_2, \bm{\tilde{s}}_3\}$ can then be used as an alternative basis for describing the system evolution.  System states and operators are transformed between bases in the usual fashion by a matrix $P$ comprised of the $\{\bm{\tilde{s}}_i\}$, entered as column vectors. Vector $\tilde{\mathcal{M}}$ and matrix $\tilde{\Gamma}$ in the new basis are given by
     \begin{eqnarray}
\tilde{\mathcal{M}} & = & P^{-1}\mathcal{M} \nonumber \\
e^{-\tilde{\Gamma} t} & = & P^{-1} e^{-\Gamma t} P \nonumber \\
                      & = & e^{-(P^{-1}\Gamma P)t}
\label{BasisTrans}
     \end{eqnarray}
with $P$ invertible since the $\bm{\tilde{s}}_i$ are linearly independent. 

The potentially tedious process of calculating 
$e^{-\tilde{\Gamma}\,t}$ from \Eq{BasisTrans} can be bypassed, with $e^{-\tilde{\Gamma}\,t}$ deduced from the action of $\Gamma$ on its eigenvectors (see Appendix \ref{Vector Model}).  In terms of constants
     \begin{equation}
\tilde s_1 = -(\bar{R} - z_1)  \quad \mathrm{and} 
             \quad \tilde s_{23} = -(\bar{R} + z_1/2),
\label{OneRealEigval}
     \end{equation}
the solution 
$\tilde{\mathcal{M}}(t) = e^{-\tilde{\Gamma} t}\tilde{\mathcal{M}}(0)$ for the time dependence of state vector 
$\tilde{\mathcal{M}}$ in the new basis is found to be 
     \begin{eqnarray}
\tilde{\mathcal{M}}(t) & = & 
  \left(\begin{array}{ccc}
     e^{\tilde s_1 t} & 0 & 0     \\
     0 & e^{\tilde s_{23} t} &  0 \\
     0 & 0 &  e^{\tilde s_{23} t}
         \end{array}
  \right) \, \times  \nonumber \\
 & & \qquad\  \left(\begin{array}{ccc}
     1 & 0 & 0     \\
     0 & \cos\varpi t & \sin\varpi t \\
     0 & -\sin\varpi t &  \cos\varpi t 
         \end{array}
  \right) \, \tilde{\mathcal{M}}(0)
\label{RotPictSoln}
     \end{eqnarray}

Viewed in the $\{\bm{\tilde{s}}_i\}$ coordinate system, the component of $\mathcal{M}$ along $\bm{\tilde{s}}_1$ (i.e., $\tilde{\mathcal{M}}_1$) decays at the rate 
$\bar{R} - z_1$, while components in the 
$(\bm{\tilde{s}}_2, \bm{\tilde{s}}_3)$-plane rotate in the plane and decay at the rate $\bar{R} + z_1/2$.
Thus, even in the most general case of three unequal rates $R_1, R_2, R_3$,
there emerges a single ``planar'' relaxation rate $R_{2s}$ and a new ``longitudinal'' relaxation rate $R_{1s}$ defined as
     \begin{equation}
R_{1s} = |s_1|= 1/T_{1s}  \quad \mathrm{and} \quad R_{2s} = |s_{23}| =  1/T_{2s}.
\label{NewLongTransRlx}
     \end{equation}

Defining $\tilde{\mathcal{M}}(t)$ as the state $\bm{M}(t) - \bm{M}_\infty$ expressed in the $\{\bm{\tilde{s}}_i\}$ coordinates and working backwards from \Eq{RotPictSoln} gives the Bloch equation in this basis as
     \begin{eqnarray}
\frac{d}{dt}\,\tilde{\mathcal{M}}(t) + 
              \tilde\Gamma\, \tilde{\mathcal{M}}(t) = 0  \nonumber \\
\nonumber \\
   \tilde\Gamma =    \left( \begin{array}{ccc}
                       R_{1s} & 0 & 0  \\
                       0 & R_{2s} & \varpi   \\
                       0 & -\varpi & R_{2s}
                    \end{array}
             \right)
\label{BlochEqNewBasis}
     \end{eqnarray}
The diagonal matrix consisting of the relaxation rates $R_{is}$ commutes with the matrix of off-diagonal elements, which generates a rotation about $\bm{\tilde{s}}_1$, and one immediately obtains the solution given in \Eq{RotPictSoln}.

One therefore has considerable latitude in the choice of $\bm{\tilde{s}}_2$ and $\bm{\tilde{s}}_3$, since all components in the plane they define decay at the same rate.  Rotating these coordinate axes in the plane by any angle results in an equally valid set of axes for representing the dynamics.  The vectors $\bm{\tilde{s}}_2$ and $\bm{\tilde{s}}_3$ constructed from a particular column in the coefficient matrices of \Eq{A0,A1} are related by such a rotation to the axes constructed from one of the other columns (excepting when one of the columns returns the irrelevant zero vector).  By contrast, $\bm{\tilde{s}}_1$ defines the unique axis for longitudinal decay, so the $\bm{\tilde{s}}_1$ chosen from different columns must be related by a scale factor.  

Note also that the rotation in the plane is \textit{not} at a constant angular frequency $\varpi$ unless $\bm{\tilde{s}}_2$ and $\bm{\tilde{s}}_3$ are orthogonal.  A component aligned with $\bm{\tilde{s}}_2$ rotates to $\bm{\tilde{s}}_3$ during a time defined by  the condition $\varpi t = \pi/2$, then rotates from there to $-\bm{\tilde{s}}_2$ in the same time.  In an oblique coordinate system, the rotations are through different angles in the same time, so clearly the angular frequency of the rotation is not constant.  

\subsubsection{\bf{Three real roots}}

In this case, all the eigenvectors are real and the new basis is simply the eigenbasis
$\{\bm{s}_1, \bm{s}_2, \bm{s}_3\}$ obtained from the roots
     \begin{equation}
s_i = -(\bar{R} - z_i)
\label{AllRealEigval}
     \end{equation}
defined in \Eq{s_i}.  The real roots $z_i$ are obtained for $\varpi^2 < 0$ in \Eq{Discr}.  Substituting $\varpi \rightarrow i\mu$ in \Eq{CubicRoots_z1} gives $z_{2,3} = -1/2\, z_1 \mp \mu$.

The matrix $\Gamma$ is obviously diagonal in its eigenbasis, and, by extension, so is the propagator in this basis.  Thus
     \begin{eqnarray}
\tilde{\mathcal{M}}(t) & = & 
  \left(\begin{array}{ccc}
     e^{s_1 t} & 0 & 0     \\
     0 & e^{s_2 t} &  0 \\
     0 & 0 &  e^{s_3 t} 
         \end{array}
  \right) \,  \tilde{\mathcal{M}}(0)
\label{RotPictSoln3real}
     \end{eqnarray}
Each component of $\mathcal{M}$ along $\bm{\tilde{s}}_i$ decays at the rate determined by $s_i$.  In contradistinction to the rates that emerge from the oscillatory solutions, here, even in the typical case of equal transverse rates $R_1 = R_2$ and longitudinal rate $R_3$, we find three distinct rates
     \begin{equation}
R_{is} = |s_i|= 1/T_{is}
\label{3RlxRates}
     \end{equation}
due to the coupling of the field with the relaxation processes.  

Given $e^{-\tilde{\Gamma} t}$ as obtained in \Eq{RotPictSoln} or (\ref{RotPictSoln3real}), the propagator in the standard coordinate basis is
$e^{-\Gamma t} = P e^{-\tilde{\Gamma} t} P^{-1} $ from \Eq{BasisTrans}. One obtains a simple, factored solution for the propagator and a correspondingly simple physical interpretation of the dynamics, with oscillation frequencies and decay rates hinging upon the primary real root $z_1$.  The dependence of this root on the fields and relaxation rates has been shown previously in \Fig{z1_w}.

\subsubsection{\bf{Degenerate roots}}
The vector model approach to obtaining the propagator is only applicable to the case of distinct eigenvalues.  Degenerate eigenvalues do not give the linearly independent eigenvectors necessary to define a new coordinate system.  However, the degeneracies are a relatively trivial component of the parameter space, at least for $R_1 = R_2$, as shown in \Fig{RootsDomain}.  Moreover, the solution has to be continuous as the degeneracies are approached, with a smooth transition from oscillatory, decaying solutions to pure decay as one crosses the parameter-space boundary identifying the degenerate solutions.

\subsection{Discussion and representative examples}

The solutions of section \ref{PropSolns} are represented in the standard coordinate system, expressed in general form for the case of three unequal relaxation rates.  Here, they are applied to specific physical examples, with $R_1 = R_2$.  The trajectories of initial states under the action of the propagator are plotted to illustrate the underlying simplicity of the dynamics and corroborate the alternative coordinate system that defines the vector model.  Parameters for the examples are chosen to demonstrate the damping and rotation that are characteristic of the dynamics for all but a small region of the parameter space.  A purely damped solution and model dynamics given by \Eq{RotPictSoln3real} is rather featureless, by comparison.  Unless stated otherwise, the first column of $\mathrm{adj}\,A_\p$ is chosen to calculate the coordinate basis $\{\bm{\tilde s}_i\}$. 

\subsubsection{\bf{Free precession, $\bm{\omega}_e = (0,0,\omega_3)$}}

When the only field in the rotating frame is the offset from resonance, $\omega_3$, the matrix $\Gamma_\p$ is the sum of a diagonal relaxation matrix and the matrix which generates a rotation about $\omega_3$.  Since they commute, the propagator factors into the product of exponential decay and a rotation, leading to the standard interpretation of the dynamics discussed previously.  This example also provides a simple context for applying the more general vector model.  The eigenvalues are easily obtained as 
$z_1 = 2R_\delta$ and $z_\pm = -R_\delta \pm i \omega_3$.  Then \Eq{RealBasis_zi} gives, upon identifying $\varpi \equiv \omega_3$ and eliminating common factors in individual columns,
     \begin{eqnarray}
\bm{\tilde s}_1 & \leftarrow & \left( \begin{array}{ccc}
   0 & 0 & 0 \\
   0 & 0 & 0 \\
   0 & 0 & 1
                     \end{array}
             \right) \qquad
\bm{\tilde s}_2  \leftarrow \left( \begin{array}{ccc}
   \omega_3 & -3R_\delta & 0 \\
   3R_\delta & \omega_3 & 0 \\
   0 & 0 & 0
                     \end{array}
             \right) \nonumber \\
\bm{\tilde s}_3 & \leftarrow & \left( \begin{array}{ccc}
   3R_\delta & \omega_3 & 0 \\
   -\omega_3 & 3R_\delta & 0 \\
    0 & 0 & 0
                     \end{array}
             \right).
     \end{eqnarray}
As noted earlier, there is always only one unique nonzero result for $\bm{\tilde s}_1$, with any apparent differences between columns simply a matter of scale.  The nonzero columns for $\bm{\tilde s}_2$ are orthogonal, as are those of $\bm{\tilde s}_3$.  The columns thus differ, as expected, by a rotation in the
$(\bm{\tilde s}_2, \bm{\tilde s}_3)$-plane, in this case by $90^\circ$.  Choosing the second column and a left-handed rotation by 
$\phi = \tan^{-1}(3R_\delta/\omega_3)$ or the first column and a right-handed rotation by $90-\phi$ gives the more typical result
$\bm{\tilde s}_2 = (0,1,0)$ and $\bm{\tilde s}_3 = (1,0,0)$ depicted in \Fig{VMFigs}a.  
The model dynamics for an initial state $\mathcal{M}_0$ is a spiral about $\bm{\omega}_e$, which is aligned along the $z$-axis, with rotation at constant angular frequency $\omega_e$ in the $(x,y)$-plane, as required.  The relaxation rate obtained from \Eq{DecayRates} or \Eq{OneRealEigval} for $z_1 = 2R_\delta$, with $\lambda_z = 2$, is $R_{1s} = R_3$, while the roots $z_\pm$ with $\lambda_z = -1$ give $R_{2s} = R_2$.

\subsubsection{\bf{On resonance, $\bm{\omega}_e = (\omega_1,\omega_2, 0)$}}
\label{Onres}
On resonance, the root $z_1 = -R_\delta$, and 
$\varpi^2 = \omega_e^2 - (3/2 R_\delta)^2$ from \Eq{params:OnRes}.  The associated eigenvector $\bm{\tilde s}_1$ is obtained by inspection from \Eq{EigVecMatrix}, with $\bm{\tilde s}_2$ and $\bm{\tilde s}_3$ obtained from
Eqs.~(\ref{RealBasis_zi}) and (\ref{A0,A1}), giving
     \begin{equation}
\bm{\tilde s}_1 = \left( \begin{array}{c}
     \omega_1 \\ \omega_2 \\ 0
                         \end{array} \right)  \qquad
\bm{\tilde s}_2 = \left( \begin{array}{c}
     -\omega_2 \\ \omega_1 \\ -\frac{3}{2}R_\delta
                         \end{array} \right)  \qquad
\bm{\tilde s}_3 = \left( \begin{array}{c}
     0 \\ 0 \\ 1
                         \end{array} \right).
\label{OnResCoordSys}
     \end{equation}

Thus, on resonance, the propagator still generates a spiral about the effective field $\bm{\omega}_e = \bm{\tilde s}_1$ with precession in the 
$(\bm{\tilde s}_2, \bm{\tilde s}_3)$-plane 
orthogonal to $\bm{\tilde s}_1$.  However, as considered in section \ref{1R2C roots},
the rotation frequency driven by $\varpi$ is not constant, since 
$\bm{\tilde s}_2$ is not perpendicular to $\bm{\tilde s}_3$.  The deviation from orthogonality, determined by the third component of $\bm{\tilde s}_2$, is small for fields that are large compared to $R_\delta$.
The respective decay rates $R_{1s}$ and $R_{2s}$ are
$R_2$ and $1/2(R_2 + R_3)$, using $\lambda_z = -1$ and $\lambda_z = 1/2$ as determined from $z_1$ and $-z_1/2$.  Components along $\bm{\tilde s}_1$, i.e., in the $(x,y)$-plane, decay at the usual spin-spin relaxation rate, as would be expected.  Components rotating in the plane orthogonal to $\bm{\tilde s}_1$ experience equal influence, on average, from their projection onto the longitudinal $z$-axis defining $\omega_3$ and their projection into the $(x,y)$-plane, so one might predict from the model that they decay at the average of the usual spin-spin and longitudinal relaxation rates.  These values for the decay rates have been obtained previously as elements of the solution in the standard coordinate system \cite{Torrey} without the physical interpretation presented here.

The trajectory for an initial state ${\cal M}_0$ due to the action of propagator $e^{-\Gamma t}$ with $\bm{\omega}_e = (\omega_1,0,0)$ and nonzero relaxation is shown in \Fig{VMFigs}b.  Values of the parameters are given in the caption. For nonzero $\omega_2$, the figure is simply rotated about the $z$-axis by angle 
$\phi = \tan^{-1}(\omega_2/\omega_1)$. The state ${\cal M}_0$ has been chosen with equal components parallel and orthogonal to $\bm{\omega}_e$ to most clearly illustrate the dynamics predicted by the vector model. The slight misalignment between $\bm{\tilde s}_2$ and the $y$-axis is evident in the figure and becomes more prominent as the magnitude of the field, $\omega_{12}$, is reduced relative to $R_\delta$.

\subsubsection{\bf{Off resonance, general $\bm{\omega}_e$}}

Most generally, $\bm{\tilde s}_1$ is not aligned with $\bm{\omega}_e$.
Dividing column $j$ of the matrix in \Eq{EigVecMatrix} by (nonzero) $\omega_j$ quantifies the degree to which $\bm{\tilde s}_1$ deviates from $\bm{\omega}_e$, due to the coupling between the fields and the relaxation. The result is an expression of the form $\bm{s}_1 = \bm{\omega}_e + \delta\bm{v}$, where vector $\delta\bm{v}$ is comprised of the second term in each row of the $j^{\mathrm{th}}$ column divided by $\omega_j$.

\begin{figure*}[h]
\includegraphics[trim = 1.75in 0 0 0, scale=.9]{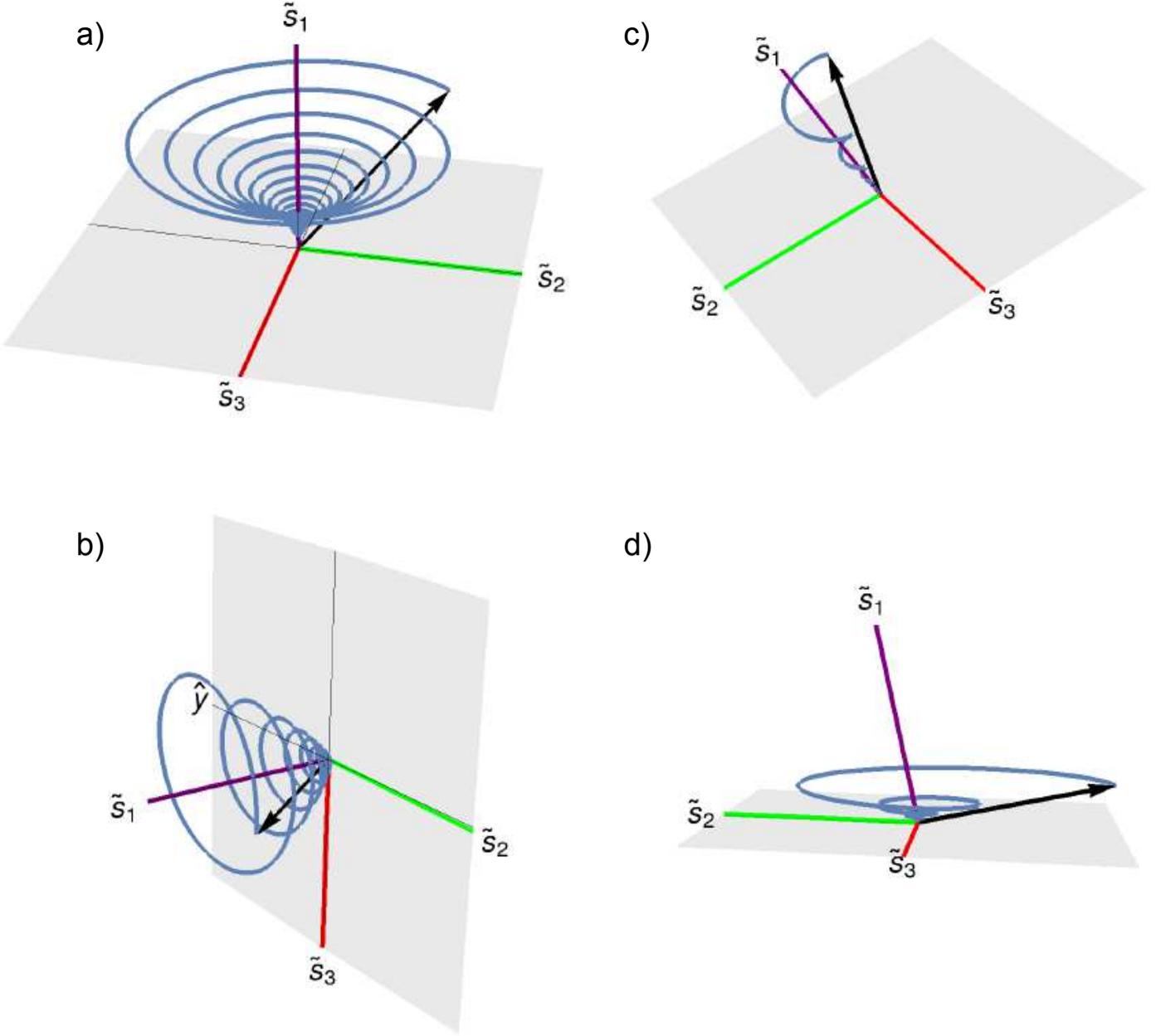}
\caption{Trajectories for initial vector $\mathcal{M}_0$ acted upon by  propagator $e^{-\Gamma t}$ are displayed in the 
$\{\bm{\tilde{s}}_1, \bm{\tilde{s}}_2, \bm{\tilde{s}}_3\}$-coordinates developed as the natural system for describing propagator dynamics. The component of $\mathcal{M}_0$ along $\bm{\tilde{s}}_1$ decays at the rate 
$\bar{R} - z_1$, while components in the 
$(\bm{\tilde{s}}_2, \bm{\tilde{s}}_3)$-plane rotate in the plane and decay at the rate $\bar{R} + z_1/2$. The different panels represent different $\mathcal{M}_0$, fields $\bm{\omega}_e$, transverse relaxation rate $R_2$, and longitudinal relaxation rate $R_3$, with details of the predicted system evolution described in more detail in the text. Physical parameters are in units inverse seconds.  
                  $\bm{(a)}$ Initial state $\mathcal{M}_0 = (-1,1,1)$.
Physical parameters $\bm{\omega}_e = (0,0,10^4)$, $R_2 = 400$, $R_3 = 200$ give coordinates $\bm{\tilde{s}}_1 = \bm{\hat z}$, 
$\bm{\tilde{s}}_2 = \bm{\hat y}$, $\bm{\tilde{s}}_3 = \bm{\hat x}$ and the well-known rotation about $\bm{\omega}_e = \omega_3$ followed by longitudinal and transverse relaxation.
     $\bm{(b)}$ Initial state $\mathcal{M}_0 = (1,-1,0)$.  
Parameters 
$\bm{\omega}_e = (5000,0,0)$, $R_2 = 400$, $R_3 = 200$ lead to coordinates $\bm{\tilde{s}}_1 = \bm{\hat x}$, 
$\bm{\tilde{s}}_2 = (0,-1,.02)$, $\bm{\tilde{s}}_3 = \bm{\hat z}$.  Rotation is also about $\bm{\omega}_e$ for $\omega_3 = 0$ (on resonance), but now $\bm{\tilde{s}}_2$ is not perpendicular to $\bm{\tilde{s}}_3$, so the rotation in the plane transverse to $\bm{\tilde{s}}_1$ is not at constant angular frequency.
        $\bm{(c)}$ Parameters 
$\bm{\omega}_e = (0,300,300)$, $R_2 = 100$, $R_3 = 1$ lead to non-orthogonal oblique coordinates $\bm{\tilde{s}}_1 = (0.12,0.69,0,71)$, 
$\bm{\tilde{s}}_2 = (0.99, 0.04, 0.12)$, 
$\bm{\tilde{s}}_3 = (0., 0.72, -0.70)$. 
Initial $\mathcal{M}_0 = (-0.12,0.69,0,71)$ 
is normal to the $(\bm{\tilde s}_2, \bm{\tilde s}_3)$-plane, but has components in the plane and along $\bm{\tilde s}_1$ in the oblique coordinate system, so spirals about $\bm{\tilde s}_1$ as shown.
        $\bm{(d)}$ Initial $\mathcal{M} = (-0.99,0.17,0)$ 
is orthogonal to $\bm{\tilde s}_1$.  Parameters
$\bm{\omega}_e = (0,3000,3000)$, $R_2 = 1000$, $R_3 = 1$ lead to nearly identical coordinates as in (c). $\mathcal{M}_0$ projects onto
$\bm{\tilde s}_1$ in oblique coordinates and therefore decays along this direction, resulting in the spiral as shown. 
}
\label{VMFigs}
\end{figure*}

In addition, $\bm{\tilde s}_1$ is typically not orthogonal to the 
$(\bm{\tilde s}_2, \bm{\tilde s}_3)$-plane.  One then has to further modify intuitions developed from orthogonal coordinate systems.  For example, in \Fig{VMFigs}c, ${\cal M}_0$  is aligned with the normal to the $(\bm{\tilde s}_2, \bm{\tilde s}_3)$-plane. It therefore has no orthogonal projection in the plane and might naively be expected to have no evolution in the plane.  However, $\bm{\tilde s}_1$ is distinctly different than the normal, and ${\cal M}_0$ is the vector sum of a component along $\bm{\tilde s}_1$ and a component parallel to the plane, which are the quantities relevant for the vector model.  As shown in the figure, the parallel component rotates and decays in the plane while the component along $\bm{\tilde s}_1$ strictly decays.  Similarly, $\mathcal{M}_0$ orthogonal to $\bm{\tilde s}_1$ as in \Fig{VMFigs}d nonetheless has a component along $\bm{\tilde s}_1$ in the oblique coordinates that decays to generate the spiral shown in the figure.  

By constrast, the dynamics viewed in standard coordinates is oscillation of each component ${\cal M}_i(t)$ combined with relaxation at two separate rates.  As in simpler examples, it can be decoupled into two independent dynamical systems, one of which rotates in a plane and decays at one rate and another which decays along a fixed axis, albeit in an oblique coordinate system.  

The deviation of $\bm{\tilde s}_1$ from the normal to the plane is quantified in Appendix \ref{Vector Model} off resonance for $\bm{\omega}_{12}$ of either $x$- or $y$-phase and the case $\omega_1=\omega_2=\omega_3$.  

\subsection{Alternative vector model}

The Bloch equation is typically represented in vector form but can be conveniently packaged in matrix form, which is the approach taken here.  The physics of its solution---the torque on a magnetic moment in a magnetic field subject to relaxation of the magnetization---can be made more explicit by returning to the original vector operations, motivated by the treatment in \cite{Jaynes} for the rotation of a vector about the field. 

Partition $\Gamma_p$ into its diagonal elements $R_{ip}$ and off-diagonal $\omega_i$, writing $\Gamma_p = \mathcal{R}_\p + \Omega$.  The diagonal matrix $\mathcal{R}_\p$ scales each component $\mathcal{M}_i$ of a vector $\mathcal{M}$ by $R_{ip}$, and $\Omega$ implements the cross product $(-\bm{\omega}_e \times \ )$.  According to \Eq{MatExpC-H}, the propagator acting on $\mathcal{M}$ generates three separate vectors $\bm{v}_n = \Gamma_p^n \mathcal{M}, (n=0,1,2)$, which can be represented starting with $\bm{v}_0 = \mathcal{M}$ as
     \begin{eqnarray}
\Gamma_\p\, \mathcal{M} &=& (\mathcal{R}_\p + \Omega )\,\bm{v}_0  \nonumber \\
  &=& (\mathcal{R}_\p\, \mathcal{M}) -\,(\bm{\omega}_e \times \mathcal{M}) \nonumber \\
  &=& \bm{v}_1  \nonumber \\
  & &  \nonumber \\
\Gamma_\p^2\, \mathcal{M}  &=& (\mathcal{R}_\p\, + \Omega)\, \bm{v}_1  \nonumber \\
  &=& (\mathcal{R}_\p^2\,\mathcal{M}) - \,
      \mathcal{R}_\p\,(\,\bm{\omega}_e \times \mathcal{M}) -\,
      \bm{\omega}_e \times (\mathcal{R}_\p \mathcal{M}) + \nonumber \\
  & & \  \bm{\omega}_e \times (\bm{\omega}_e \times \mathcal{M}) \nonumber \\
  &=& (\mathcal{R}_\p^2\,\mathcal{M}) - 
      \mathcal{R}_\p\,(\,\bm{\omega}_e \times \mathcal{M}) -\,
      \bm{\omega}_e \times (\mathcal{R}_\p \mathcal{M}) + \nonumber \\
  & & \ \bm{\omega}_e\,(\bm{\omega}_e \cdot \mathcal{M}) - 
               \omega_e^2\,\mathcal{M}  \nonumber \\
  &=& \bm{v}_2
\label{GammaM}
     \end{eqnarray}
Each succeeding $\bm{v}_n$ is a nonuniform scaling of the previous $\bm{v}_{n-1}$ added to a vector ($\bm{v}_{n-1} \times \bm{\omega}_3$) that is orthogonal to $\bm{v}_{n-1}$.
The time dependence of $\bm{v}_n$ is given by the associated term $a_n(t) e^{-\bar{R} t}$ found in Eqs.~(\ref{MatExp1}--\ref{MatExp3}).  The $a_n(t)$ are factored as the product of a matrix $W(z_1)$ and vector $\bm{u}(t)$. Each $a_n(t)$ is merely a different linear combination of the same three simple functions $u_i(t)$ that comprise the components of $\bm{u}$, weighted according to the corresponding elements from row $n$ of the matrix $W$.  A given $\bm{v}_n(t)$ thus maintains a fixed orientation, changing length with a time dependence consisting of the different weightings of the $u_i(t)$ for different $\bm{v}_n$.  The trajectory $\mathcal{M}(t) = \sum_n \bm{v}_n(t)$ can thus be represented in terms of the decaying oscillations of three vectors fixed in place.

Alternatively, expand 
$(\Id, \Gamma_p, \Gamma_p^2)\,W(z_1)\bm{u}(t)$ and group terms of the same time dependence $u_i(t)$, as, for example, in \Eq{AppSumMatExp1}.  The propagator applied to $\mathcal{M}$ gives 
three different linear combinations of the $\bm{v}_n$, with a time dependence $u_i(t)$ for the $i^{\mathrm{th}}$ combination.  The resulting interpretation of $\mathcal{M}(t)$ is similar to the previous paragraph, but the functional form of the decaying oscillations is simpler using this different set of vectors.

\subsection{The Bloch equation as a system of coupled oscillators}

Any quantum N-level system can be represented as a system of coupled harmonic oscillators \cite{QM-SHO}, albeit requiring negative or even antisymmetric couplings.  The Bloch equation is perhaps particularly interesting, since it incorporates dissipation for the most elementary case, i.e., 2-level systems.  

Expressing \Eq{BlochEq} in terms of 
$\mathcal{M}(t) = \bm{M}(t) - \bm{M}_\infty$ yields a homogeneous first-order differential equation and an alternative route to the solution,
\Eq{BlochSol:2}.  Differentiating again with respect to time and substituting $\mathcal{\dot M} = -\Gamma\mathcal{M}$ gives
     \begin{equation}
\mathcal{\ddot M}(t) = \Gamma^2 \mathcal{M}(t),
\label{OscEq}
     \end{equation}
with
\begin{widetext}
     \begin{equation}
\Gamma^2 = \left[
                 \begin{array}{ccc}
  -(\omega _2^2 + \omega _3^2) + R_1^2  & 
   \ \ \ \omega_1 \omega_2 + \omega _3(R_1 + R_2) & 
   \ \ \ \omega_1 \omega_3 - \omega _2(R_1 + R_3) 
\\
   \omega _1 \omega _2 - \omega _3(R_1 + R_2) &
   \ \ \ -(\omega _1^2 + \omega _3^2) + R_2^2 & 
   \ \ \ \omega_2 \omega_3 + \omega _1(R_2 + R_3)
\\
   \omega_1 \omega_3 + \omega _2(R_1 + R_3) &
   \ \ \ \omega_2 \omega_3 - \omega _1(R_2 + R_3) &
  \ \ \ -(\omega_1^2 + \omega_2^2 ) + R_3^2 
\\
                 \end{array}
           \right].
\label{Gamma sq}
     \end{equation}
\end{widetext}

\begin{figure*}[h]
\includegraphics[scale=1]{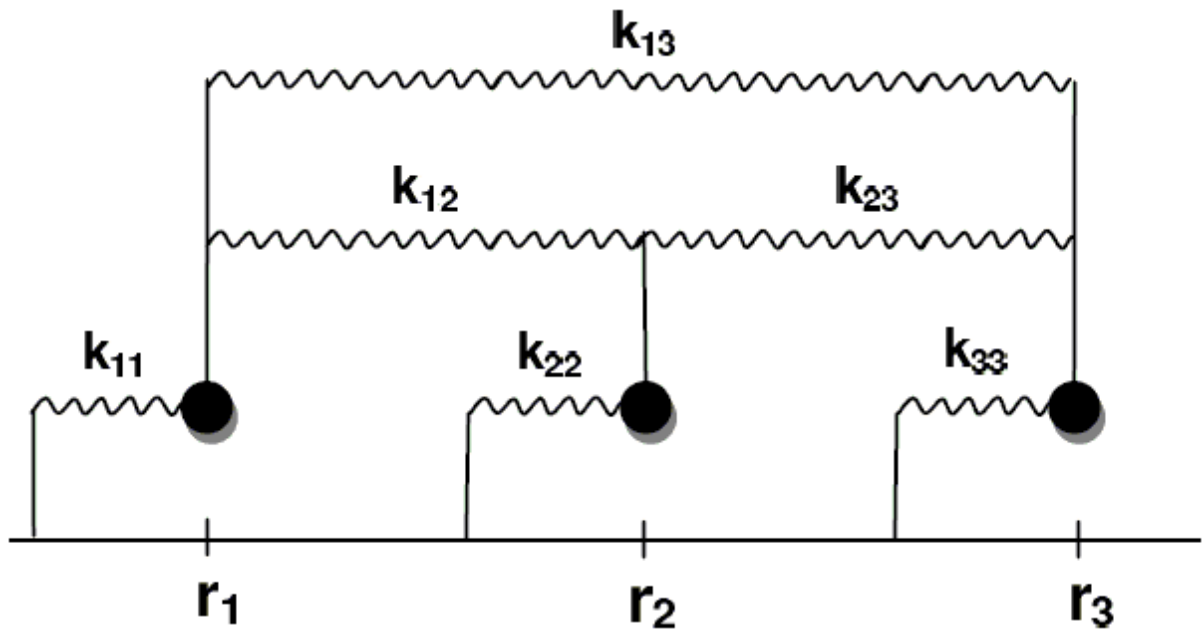}
\caption{The Bloch equation is shown in the text to model the displacements, from equilibrium positions $r_i = 0$, of an ideal frictionless system of three unit masses coupled by springs of stiffness $k_{i j}$.
}
\label{CoupledMasses}
\end{figure*}

As considered previously \cite{QM-SHO}, damping is provided by the antisymmetric part of $\Gamma^2$ in addition to the terms $R_i^2$ on the diagonal.  For the system of three coupled oscillators illustrated in \Fig{CoupledMasses}, the displacement $r_i$ of mass $m_i$ from equilibrium is equal to $\mathcal{M}_i$.  We can write the coupling constants $k_{ij} = \kappa_{ij} + \sigma_{ij}$ in terms of symmetric $\kappa_{ij}$ and anti-symmetric $\sigma_{ij}$ connected in parallel.  Then, by inspection, $\kappa_{ij} = \omega_i\omega_j$ and
$\sigma_{ij} = \varepsilon_{ijk}(R_i + R_j)\omega_k$, assuming unit masses, and $k_{ii} = -(\Gamma^2)_{ii} - \sum_j k_{ij}$.  
For a given positive $\sigma_{ij}$, a positive displacement of mass $m_j$ results in a positive force on $m_i$.  The resulting positive displacement of $m_i$ provides a negative force on $m_j$ due to $\sigma_{ji} < 0$ which opposes the original displacement of $m_j$ and damps the motion.  Stated differently, energy transferred from $m_j$ to $m_i$ is not reciprocally transferred back from $m_i$ to $m_j$, and the motion is quenched.  An antisymmetric coupling acts as a negative feedback mechanism that curbs system oscillations.

The usual representation of damped oscillators employs a velocity-dependent friction force. The above implementation is frictionless.  It provides an alternative model for investigating dissipative processes with the potential for new insights within the well understood context of coupled harmonic oscillations.

\section{Conclusion}
A complete solution of the Bloch equation has been presented together with intuitive visual models of its dynamics.  The solution is valid for arbitrary system parameters, yet is simpler than previous solutions. 
It can be expressed as the product of three separate terms:  one which depends directly on the physical parameters of the problem through the matrix $\Gamma_p$, a term that depends on the roots of a cubic characteristic polynomial for the problem, and a term that gives the time dependence, which in turn is solely a function of the roots.  Moreover, the time evolution of the system as a function of the physical parameters has been made more explicit and apparent.  

The solutions depend critically on the three polynomial roots.  Quantitative relations have been derived for the physical parameters that define the possible system dynamics: (i) oscillatory, underdamped evolution for one real and two complex-conjugate roots, (ii) non-oscillatory, overdamped evolution for three real roots, and (iii) non-oscillatory, critically damped evolution for doubly or triply degenerate (real) roots.  The damping rates and the frequency driving the oscillatory behavior have been reduced to simple functions of a single root which is obtained as a straightforward function of the system parameters.  In addition, a linear relation has been derived for the system parameters as a function of this real root, which provides a straightforward graphical realization of the damping rates and frequency for a given physical  configuration. 

An intuitive dynamical model developed here transforms the Bloch equation to a frame in which damping commutes with a rotation, providing a propagator for the time evolution of the system that is the product of a rotation times a decay, in either order. The decay rates in this frame result from interaction/coupling of the fields with the spin-lattice and spin-spin relaxation processes. The model was motivated by well-known visual models for simple cases such as equal relaxation rates or free precession (no fields transverse to the longitudinal, $z$-axis).  The system state in such cases rotates about the effective field, with concurrent exponential decay of the longitudinal and transverse components.  The extended model retains the same essential features: rotation, exponential decay of the invariant component in the rotation analogous to longitudinal relaxation, and a separate decay of the rotating components analogous to transverse relaxation.  
An alternative vector model has also been provided, as well as a representation of the Bloch equation as a system of coupled harmonic oscillators.  The net result of the solutions and models is more direct physical insight into the dynamics of the Bloch equation.

\begin{acknowledgments}
The author gratefully acknowledges support from the National Science Foundation under Grant CHE-1214006.
\end{acknowledgments}


\begin{widetext}
\par\vfill\eject
\end{widetext}
\appendix
\section{Cubic Polynomials with Real Coefficients}
\label{AppPolyRoots}
The standard solutions for the three roots of \Eq{SCForm}, cast here in terms of 
     \begin{equation}
\Lambda_\pm = \big[-b/\,2 \pm \sqrt{(b/2)^2 + (a/3)^3}\;\big]^{1/3},
\label{Lambda_pm}
     \end{equation}
are
     \begin{eqnarray}
z & = & \bigg\{\Lambda_+ + \Lambda_-, -\frac{\Lambda_+ + \Lambda_-}{2}\pm \sqrt{-3}\;\frac{\Lambda_+ - \Lambda_-}{2} \bigg\}, \nonumber \\
  & = & \{z_1, z_\pm\}.
\label{CubicRoots}
     \end{eqnarray}
These solutions can be consolidated in a convenient form that does not appear to have been employed heretofore.  Substituting 
$(\Lambda_+ - \Lambda_-) = \big[(\Lambda_+ + \Lambda_-)^2 - 4\Lambda_+\Lambda_-\big]^{1/2}$ and noting
$\Lambda_+\Lambda_- = -a/3$ gives
     \begin{eqnarray}
z_1 & = & \Lambda_+ + \Lambda_- \nonumber \\
z_\pm  & = & -\frac{1}{2}z_1\, \pm \, 
        i \sqrt{3}\,\sqrt{\, \left(\frac{z_1}{2}\right)^2 +\frac{a}{3}\,} \nonumber \\
       & = & -\frac{1}{2}z_1\, \pm \,i\,\varpi
\label{AppCubicRoots_z1}
     \end{eqnarray}
in terms of a discriminant 
     \begin{equation}
\varpi^2 = 3\big[(z_1/2)^2 + a/3\big].
\label{AppDiscr}
     \end{equation}
Any polynomial with real coefficients has at least one real root. Therefore $\varpi^2>0$ gives one real and two complex conjugate roots, with three real roots resulting from $\varpi^2 \le 0$. 

One can then employ simple forms for $z_1$ \cite{MiuraCubicRoots, McKelveyCubicRoots}. The number of conditional dependencies relating the cited expressions for $z_1$
to the signs and relative magnitudes of $a$ and $b$ can be further simplified in terms of
     \begin{eqnarray}
\alpha & = & |\,a/3\,| \nonumber \\
\beta & = & |\,b/2\,|  \nonumber \\
\gamma & = & \dfrac{\beta}{\alpha^{3/2}}.
\label{AlphaBetaGamma}
     \end{eqnarray}
Then the roots can be calculated according to their domain of applicability as
\begin{subequations}
\label{Root_z1}
     \begin{eqnarray}
a > 0 \phantom{jjjjj} & & \nonumber \\      
  & & \varphi \equiv \tfrac{1}{3}\sinh^{-1}\gamma  \nonumber \\
  & & x_1 \equiv \mbox{sgn}(b)\sinh\,\varphi  \nonumber \\ 
  & & z_1 = -2\,\sqrt{\alpha}\,x_1
\label{Root_z1:1}   \\
  & & \quad\ 
\varpi = \sqrt{3\alpha(x_1^2 + 1)} = \sqrt{3\alpha}\cosh\varphi \nonumber \\
  & & z_\pm = \sqrt{\alpha}x_1 \pm i\,\varpi \nonumber \\
a < 0 \phantom{jjjjj} & &  \nonumber \\  
\gamma \ge 1 & & \phantom{jjj}\phantom{(b/2)^2 + (a/3)^3 > 0} \nonumber \\
  & & \varphi \equiv \tfrac{1}{3}\cosh^{-1}\gamma  \nonumber \\
  & & x_1 \equiv \mbox{sgn}(b)\cosh\,\varphi  \nonumber \\ 
  & & z_1 = -2\,\sqrt{\alpha}x_1
\label{Root_z1:2}  \\
  & & \quad\ 
\varpi = \sqrt{3\alpha(x_1^2 - 1)} = \sqrt{3\alpha}\sinh\varphi \nonumber \\
  & & z_\pm = \sqrt{\alpha}x_1 \pm i\,\varpi  \nonumber \\
  & & \phantom{z_\pm} \rightarrow \sqrt{\alpha}x_1 \qquad \gamma = 1
        \nonumber  \\
\gamma \le 1 & & \phantom{jjj}\phantom{(b/2)^2 + (a/3)^3 > 0} \nonumber \\
  & & \varphi \equiv \tfrac{1}{3}\cos^{-1}\gamma  \nonumber \\
  & & x_1 \equiv \mbox{sgn}(b)\cos\,\varphi  \nonumber \\ 
  & & z_1 = -2\,\sqrt{\alpha}\,x_1
\label{Root_z1:3} \\
  & & \quad\ 
\varpi = i\,\sqrt{3\alpha(1 - x_1^2)} = i\,\sqrt{3\alpha}\sin\varphi \nonumber \\
  & & \quad\ \phantom{\varpi} = i\,\mu \nonumber \\
  & & z_\pm = \sqrt{\alpha}x_1 \pm \mu \qquad \mathrm{or,\  alternatively} \nonumber \\
  & &   \nonumber \\
  & & \varphi \equiv \tfrac{1}{3}\sin^{-1}\gamma  \nonumber \\
  & & x_1 \equiv \mbox{sgn}(b)\sin\,\varphi  \nonumber \\ 
  & & z_1 = +2\,\sqrt{\alpha}\,x_1
\label{Root_z1:3p} \\
  & & \quad\ 
\varpi = i\,\sqrt{3\alpha(1 - x_1^2)} = i\,\sqrt{3\alpha}\cos\varphi \nonumber \\
  & & \quad\ \phantom{\varpi} = i\,\mu \nonumber \\
  & & z_\pm = \sqrt{\alpha}x_1 \pm \mu  \nonumber \\
a = 0 \phantom{jjjjj} & &  \nonumber \\      
 & & z_1 = -\mbox{sgn(b)}\sqrt[3]{|b|}   
\label{Root_z1:4}  \\  
 & & z_\pm = -\frac{1}{2}z_1 (1 \pm i\,\sqrt{3}) \nonumber 
     \end{eqnarray}
\end{subequations}

For $a>0$ or $a<0$ and $ \gamma > 1$, there is one real root and complex conjugate roots $z_\pm$.  
For ${a<0,\, \gamma < 1}$, there are three real roots, with $\gamma = 1$ in Eq.\,(\ref{Root_z1:2}) or (\ref{Root_z1:3}) giving 
$\varphi = 0 = \varpi$ and two degenerate roots $z_+ = z_-$, while
(\ref{Root_z1:3p}) reorders the roots relative to (\ref{Root_z1:3}), so that the nondegenerate root for the case $\gamma = 1$ is one of the $z_\pm$.  Results for $a=0$ are straightforwardly obtained from 
Eqs.~(\ref{CubicRoots}) and (\ref{CubicRoots_z1}), or using the expressions in (\ref{Root_z1:1}) and (\ref{Root_z1:2}), with
$\sinh^{-1}\gamma \rightarrow \cosh^{-1}\gamma \rightarrow \ln(2\gamma)$ in the limit $\gamma \rightarrow \infty$.  Terms then result that are multiplied by $\sqrt{\alpha}$, cancelling the singularity at $a=0$.
For the case $a=0=b$, there are three equal roots $z_i = 0$.
\section{Calculation of $\bm{e^{-\Gamma_\p t}}$}
\label{App:CalcProp}
\subsection{First-order pole}

Consider the case of one real root $z_1$ and two complex conjugate roots
$z_{2,3} = -1/2 z_1 \pm i\,\varpi$, as given by \Eq{CubicRoots_z1}, with
$\varpi^2 = 3(z_1/2)^2 + a > 0$. 
Using \Eq{adjA_Poly} for $\mathrm{adj}A_\p$ in \Eq{MatExp} gives
     \begin{eqnarray}
e^{-\Gamma_\p t}\, &=& \sum_{i=1}^3 
       \frac{\mathrm{adj}\,A_{\p}(z_i)}{q^{\,\prime}(z_i)} e^{\,z_i t} \nonumber \\
  &=&\sum_{i=1}^3 \frac{e^{\,z_i t}}{q^{\,\prime}(z_i)}\, 
            \sum_{n=0}^2 A_{n\p} z_i^n,
\label{AppMatExp1}
     \end{eqnarray}
with $q^\prime(z_i) = \prod_{j \neq i}(z_i-z_j)$, as discussed in section \ref{Simple pole}, and $A_{2\p} \equiv \Id$.

Evaluating the $q^{\,\prime}(z_i)$ and using \Eq{Discr} for $\varpi^2$ gives
     \begin{eqnarray}
q^{\,\prime}(z_1) &=& (z_1 - z_2)(z_1 - z_3) \nonumber \\
  &=& (3/2 z_1)^2 + \varpi^2 \nonumber \\
  &=& 3z_1^2 + a, \nonumber \\
  & & \nonumber \\
q^{\,\prime}(z_2) &=& (z_2 - z_1)(z_2 - z_3) \nonumber \\
  &=& -q^{\,\prime}(z_1)(z_2 - z_3)/(z_1 - z_3) \nonumber \\
  &=& -i\,(3z_1^2 +a)\,2\varpi/(3/2\, z_1 + i\,\varpi) \nonumber \\
 & &  \nonumber \\
q^{\,\prime}(z_3) &=& [q^{\,\prime}(z_2)]^*
\label{Denom}
     \end{eqnarray}

Since $z_3 = z_2^*$, each coefficient $(A_n)_\p$ in \Eq{MatExp} multiplies a sum $S_n$
     \begin{eqnarray}
S_n &=& \sum_{i=1}^3 z_i^n \frac{e^{z_i t}}{q^{\,\prime}(z_i)}
       \nonumber \\
  &=& z_1^n \frac{e^{z_1 t}}{q^{\,\prime}(z_1)} 
    + 2\, \mathrm{Re}\bigg(z_2^n \frac{e^{z_2 t}}{q^{\,\prime}(z_2)}\bigg). 
\label{Sn}
     \end{eqnarray}
Substituting $z_2 = -z_1/2 + i\,\varpi$ gives
\begin{widetext}
     \begin{eqnarray}
S_0 &=& \frac{1}{3z_1^2 + a} \bigg\{\,e^{z_1 t} - e^{-z_1 t/2}\Big[
    \cos\varpi t + \tfrac{3}{2} z_1 \frac{\sin\varpi t}{\varpi}\,\Big]\,\bigg\}
       \nonumber \\
& & \nonumber \\
S_1 &=& \frac{1}{3z_1^2 + a} \bigg\{\,z_1 e^{z_1 t} - e^{-z_1 t/2}\Big[
    z_1 \cos\varpi t - (\tfrac{3}{4} z_1^2 + \varpi^2) \frac{\sin\varpi t}{\varpi}\,\Big]\,\bigg\}
       \nonumber \\
S_2 &=& \frac{1}{3z_1^2 + a} \bigg\{\,z_1^2 e^{z_1 t} + e^{-z_1 t/2}\Big[
   (\tfrac{5}{4} z_1^2 + \varpi^2) \cos\varpi t - 
   \tfrac{1}{2}z_1\,(\tfrac{3}{4}z_1^2 - \varpi^2)\frac{\sin\varpi t}{\varpi}\,\Big]\,\bigg\},
\label{S012}
     \end{eqnarray}
\end{widetext}

We then have, using \Eq{adjAp_coeff_Gamma} for the $(A_n)_\p$,
\begin{widetext}
     \begin{eqnarray}
e^{-\Gamma_\p t}\, &=& (A_0)_\p S_0 + (A_1)_\p S_1 + (A_2)_\p S_2   \nonumber \\
  &=& (\Gamma_p^2 + a) S_0 + (-\Gamma_p) S_1 + \Id S_2  \nonumber \\
  &=& \frac{1}{3z_1^2 + a}\,
\bigg\{\, e^{z_1 t}\,\Big[\, (z_1^2 + a)\,\Id - z_1\Gamma_p + \Gamma_p^2\,\Big] +
          e^{-z_1 t/2}\,\Big[\, 2z_1^2\,\Id + z_1\Gamma_p - \Gamma_p^2\,\Big]\,\cos\varpi t  - \nonumber \\
  & &
  \qquad\qquad\quad  
   e^{-z_1 t/2}\,\Big[\, a z_1\,\Id + (\tfrac{3}{2}z_1^2 + a)\,\Gamma_p + 
                \tfrac{3}{2}z_1\Gamma_p^2\,\Big]\,
                 \frac{\sin\varpi\, t}{\varpi}\,
\bigg\} 
\label{AppSumMatExp1}
     \end{eqnarray}
\end{widetext}

Arranging coefficients of ($\Id, -\Gamma_p ,\Gamma_p^2$) in a matrix for each time-dependent term in the solution gives the result in \Eq{MatExp1}. All three roots are real when $\varpi^2 < 0$, which is the case for $a<0$ and $\gamma < 1$. Then 
$\varpi \rightarrow i\mu$ in \Eq{MatExp1}, with
$\mu^2 = |3(z_1^2/2) + a|$ and $a = -|a|$.

\subsection{Second-order pole}
\label{App2ndOrder}
The case $\varpi=0$ gives doubly-degenerate real roots $z_2 = z_3 = -z_1/2$, and the characteristic polynomial
$q\,(z)\rightarrow (z-z_1)(z-z_2)^2$.  
The residue at $s = z_2$ in \Eq{F(t)} requires the derivative of 
$e^{s t}\mathrm{adj}\,A_\p(s)/(s-z_1)$ with respect to $s$, evaluated at $s=z_2$.  Expanding $\mathrm{adj}\,A_\p$ using Eqs.~(\ref{adjA_Poly}) and (\ref{adjAp_coeff_Gamma}) as above, utilizing a common denominator 
$(z_2 -z_1)^2$, and substituting $z_2 = -z_1/2$ gives 
     \begin{eqnarray}
\mathrm{Res}(z_2) &=& \frac{t\, e^{z_2 t} \mathrm{adj}\,A_\p(z_2)}{z_2-z_1} -
             \frac{e^{z_2 t} \mathrm{adj}\,A_\p(z_2)}{(z_2-z_1)^2} + \nonumber \\
  & & 
             \frac{e^{z_2 t} (\mathrm{adj}\,A_\p)^\prime\big|_{z_2}}{z_2-z_1}
       \nonumber \\
  &=& e^{-z_1 t/2}\bigg[\,\bigg(\frac{8}{9} + \frac{1}{3} z_1 t \bigg)\,\Id +
     \bigg(\frac{4}{9}\,z_1^{-1} - \frac{1}{3}\,t \bigg)\,\Gamma_p -     \nonumber \\
  & & \qquad\qquad
       \bigg(\frac{4}{9}\,z_1^{-2} + \frac{2}{3}\,t\,z_1^{-1} \bigg)\,\bigg]
\label{Res(z2)}
     \end{eqnarray}
The contribution from the first-order pole at 
$z_1$ is obtained as before from the $i=1$ term of \Eq{MatExp1} with 
$a = -3(z_1/2)^2$, since $\varpi = 0$, to yield the result of \Eq{MatExp2}.

\section{Existence of Degenerate Roots} 
\label{AppDegRoots}
The characteristic polynomial for the case $R_1=R_2$ has degenerate roots for $D(a,b) = 0$ (cf. \Eq{Dfunc}), which requires $a<0$. The special case $a=0=b$ discussed in section \ref{DegRoots} gives $\omega_3^2 = 1$ and $\omega_{12}^2 = 8$, normalized to $R_\delta^2/3$.  More generally, scale $\omega_3^2$ and $\omega_{12}^2$ in terms of the same normalization as 
     \begin{equation}
\omega_3^2 = \lambda_3 \;R_\delta^2/3, 
\label{w3_scale}
     \end{equation}
where $\lambda_3 \geq 0$, and 
     \begin{equation}
\omega_{12}^2 = (\eta-\lambda_3 +9/4)\;R_\delta^2/3.
\label{wa_scale}
     \end{equation}
Then $D(a,b)=0$ gives
     \begin{equation}
\eta^3 + a_\eta\eta + b_\eta = 0,
\label{etaCubic}
     \end{equation}
with
     \begin{eqnarray}
\dfrac{a_\eta}{3} & = & - \bigg(\frac{3}{2}\bigg)^4\,(8\lambda_3 + 1) \nonumber \\
\dfrac{b_\eta}{2} & = & \bigg(\frac{3}{2}\bigg)^6\,(8\lambda_3 ^2 + 20\lambda_3 - 1).
\label{ab_eta}
     \end{eqnarray}
The roots $\eta_{\,1}(\lambda_3 )$ and $\eta_\pm(\lambda_3 )$ of \Eq{etaCubic} can then be obtained using Eqs.~(\ref{Root_z1}) with the appropriate substitution of variables.  Only those solutions such that
\mbox{$\omega_{12}^2 \geq 0$} (i.e., is real) are of interest.  The results, outlined in detail below, are that (i) there are no degenerate roots if $\omega_3^2 > R_\delta^2/3$; and (ii) for each $\omega_3$ satisfying $0 \leq \omega_3^2 \leq R_\delta^2/3$, there are two values of $\omega_{12}^2$ that give degenerate roots. 

\squishlist
\item[Note for use in what follows that] 
\item[$\cdot$] $a_\eta < 0$ for all $\lambda_3 \geq 0$
   \squishlist
      \item[] $\therefore$ no \Eq{Root_z1:1} solutions for $\eta$
   \squishend
\item[$\cdot$] $\sqrt{\alpha_\eta} = \sqrt{|a_\eta/3|} =
                \frac{9}{4}\sqrt{8\lambda_3 +1}$
\item[$\cdot$] $b_\eta = 0$ for 
      $\lambda_3 \,=\, \frac{3}{4}(\sqrt{3} - \frac{5}{3}) \equiv \lambda_{b} \approx 0.05$
\item[$\cdot$] $D(a_\eta,b_\eta) = \dfrac{3^{12}}{2^{\,6}}\,\lambda_3 \,(\lambda_3 -1)^3$ 
\item[$\cdot$] $\gamma_\eta(\lambda_3) = 
                \dfrac{|8\lambda_3 ^2 + 20\lambda_3 -1|}                                     {(8\lambda_3 + 1)^{3/2} }$ \qquad (cf. \Eq{AlphaBetaGamma})
   \squishlist
      \item[] $\gamma_\eta(0) = 1,\quad \gamma_\eta(\lambda_b) = 0,
                     \quad \gamma_\eta(1) = 1$
   \squishend
\squishend

\squishlist
\item[1)] If $\lambda_3 > 1$, then
  \squishlist
  \item[$\cdot$] $D(a_\eta,b_\eta) > 0$, equivalent to $\gamma_\eta > 1$
  \item[$\cdot$] there is one real solution $\eta_{\,1}$ from \Eq{Root_z1:2}
  \item[$\cdot$] Define $\varphi_\eta = \tfrac{1}{3}\cosh^{-1}\gamma_\eta$
  \item[$\cdot$] $b_\eta > 0$
  \item[]
  \item[$\cdot$] $\eta_{\,1} = -2\sqrt{\alpha_\eta}\cosh\varphi_\eta$
     \squishlist
        \item[] $\cosh\varphi_\eta \ge 1$ for all $\varphi_\eta$,
        \item[]  $2\sqrt{\alpha_\eta} > \frac{9}{2}\,(3)$
     \squishend
  \squishend
  \item[] $\therefore \eta_{\,1} < -\frac{27}{2}$
  \item[] $\implies \omega_{12}^2 \sim (\eta_{\,1} + \frac{9}{4} - \lambda_3 )
         < -\frac{45}{4} - \lambda_3 < 0$
\item No real $\omega_{12}$ such that \Eq{SCForm} has degenerate roots for $\omega_3^2 = \lambda_3 \;R_\delta^2/3 > R_\delta^2/3$ 
\squishend

\squishlist
\item[2)] If $\lambda_3 \leq 1$, then
  \squishlist
    \item[$\cdot$] $\omega_{12}^2 \sim (\eta + \frac{9}{4} - \lambda_3 ) \ge 0$
                   for $\eta \ge 0$
    \item[$\cdot$] $D(a_\eta,b_\eta) \leq 0$, equivalent to $\gamma_\eta \leq 1$
    \item[$\cdot$] there are three real solutions $\eta_{\,1}, \eta_\pm$ from \Eq{Root_z1:3p}
    \item[$\cdot$] Define $\vartheta = \frac{1}{3}\,\sin^{-1}(\gamma_{\,\eta})$
           \squishlist
             \item[] $0 \le \sin\vartheta \le \frac{1}{2}$
           \squishend
  \squishend

  \squishlist
    \item[(a)] If $\lambda_{\,b} \leq \lambda_3 \leq 1$, then

                \quad\   $0 \le\gamma_\eta \le 1$,

                \quad\   $0 \le \vartheta \le \pi/6$, 

                \quad\ $b_\eta \geq 0$
      \squishlist
        \item[$\cdot$] $\eta_{\,1} = 2\sqrt{\alpha_\eta}\sin\vartheta$
        \item[] $\ \ \therefore \phantom{\eta_1} \eta_{\,1} \ge 0$
        \item[] $\implies \omega_{12}^2\, > 0$
        \item[]
        \item[$\cdot$] $\eta_+ = -\sqrt{\alpha_\eta}\,\sin\vartheta\, +\, 
                    \sqrt{3}\,(\alpha_\eta - \alpha_\eta\sin^2\vartheta)^{1/2}$ 
        \item[] $\phantom{\eta_+} = 2\sqrt{\alpha_\eta}\,\sin(\pi/3 - \vartheta)$
        \item[] $\ \ \therefore \phantom{\eta_1} \eta_+ \geq 0$
        \item[] $\implies \omega_{12}^2\, > 0$
      \squishend

    \item[(b)] If $0 \leq \lambda_3 \leq \lambda_{\,b}$, then

                \quad\   $1 \ge \gamma_\eta \ge 0$,

                \quad\   $\pi/6 \ge \vartheta \ge 0$, 

                \quad\  $b_\eta \leq 0$               
      \squishlist
        \item[$\cdot$] $\eta_{\,1} = -2\sqrt{\alpha_\eta}\sin\vartheta$
        \item[] $\ \ \therefore \  -\frac{9}{4} \,\leq\, \eta_{\,1} \leq\, 0$
        \item[] $\implies \omega_{12}^2 \sim \eta_{\,1} + \frac{9}{4} - \lambda_3 \geq 0$,
        \item[] \qquad\qquad since $\eta_{\,1} \in [-\frac{9}{4},0\,]$ as 
                  $\lambda_3 \in [0,\lambda_b]$
        \item[]
        \item[$\cdot$] $\eta_+ = \sqrt{\alpha_\eta}\,\sin\vartheta\, +\, 
                    \sqrt{3}\,(\alpha_\eta - \alpha_\eta\sin^2\vartheta)^{1/2}$ 
        \item[] $\phantom{\eta_+} = 2\sqrt{\alpha_\eta}\,\sin(\pi/3 + \vartheta)$
        \item[] $\therefore \eta_+ \geq 0$
        \item[] $\implies \omega_{12}^2\, > 0$
       \squishend
  \squishend
\item 2 real $\omega_{12}^2$ such that \Eq{SCForm} has degenerate roots for $0 \,\leq\, \omega_3^2 \,\leq\, R_\delta^2/3$ 

\squishend

The solutions for $\omega_{12}^2$ become equal at 
$\omega_3^2 = R_\delta^2/3$, as shown in \Fig{RootsDomain}, corresponding to the case
$a=0=b$. There is then a three-fold degenerate root $z=0$ of \Eq{SCForm}.  
Recall that a solution to $D(a,b)=0$ for real $a,b$ requires
$a = \omega_{12}^2 + \omega_3^2 - 3R_\delta^2 \le0$, which is readily verified for the solutions obtained above.  Scaling $a$ according to Eqs.~(\ref{w3_scale}) and (\ref{wa_scale}), dividing by $R_\delta^2/3$, and using the maximum value $\eta_\mathrm{max} = \sqrt{\alpha_\eta} = 27/4$ at $\lambda_3 =1$ gives
     \begin{eqnarray}
a &\sim& (\eta -\lambda_3 + \tfrac{9}{4}) + \lambda_3 - 9 \nonumber \\
  &\le& \tfrac{27}{4} + \tfrac{9}{4} - 9 = 0.
\label{a<=0}
     \end{eqnarray}
\section{Vector Model}
\label{Vector Model}

There is a simple physical interpretation for the action of the propagator $e^{-\Gamma\,t}$ when, as is most common, the matrix $\Gamma$ has three distinct eigenvalues.  Supplementary details of the model introduced in section \ref{VM} are presented here.  Consider the case of one real eigenvalue and two complex conjugate eigenvalues.   Results for the other possibility, that of three real eigenvalues, are obtained directly from \Eq{EigVecMatrix} in what follows.

The eigenvalues of $-\Gamma$ are the roots $s_1 = z_1 - \bar{R}$ and 
$s_{2,3} \equiv s_\pm =-z_1/2 \pm i\,\varpi -\bar{R}$, obtained from \Eq{s_i}, with real $z_1$ given in Eqs.~(\ref{Root_z1}). The  associated eigenvectors are $\bm{s}_1$ and the complex conjugate pair $\bm{s}_\pm$. 
The relation between $\bm{s}_\pm$ and the real vectors $\bm{\tilde{s}}_2$ and $\bm{\tilde{s}}_3$  defined in \Eq{Real s_2,3} is
     \begin{eqnarray}
\bm{\tilde{s}}_2 &=& \tfrac{1}{2}\,(\bm{s}_+ + \bm{s}_-)\qquad\qquad 
\bm{\tilde{s}}_3 = -\tfrac{i}{2}\,(\bm{s}_+ - \bm{s}_-)  \nonumber \\
\bm{s}_+ &=& \bm{\tilde{s}}_2 + i\,\bm{\tilde{s}}_3 
        \qquad\ 
\bm{s}_- = \bm{\tilde{s}}_2 - i\,\bm{\tilde{s}}_3\,.
\label{ComplexEigVec}
     \end{eqnarray}
Defining $\bm{\tilde{s}}_1 \equiv \bm{s}_1$ gives a set $\bm{\tilde{s}}_i$ of three linearly independent vectors that can be used as an alternative basis for representing arbitrary system states.  We then have
     \begin{eqnarray}
-\Gamma\, \bm{\tilde{s}}_2 &=& \tfrac{1}{2}\,(s_+ \bm{s}_+ + s_- \bm{s}_-) =
                       \tfrac{1}{2}\,(s_+ \bm{s}_+ + s_+^* \bm{s}_+^*)  \nonumber \\
& & \nonumber \\
e^{-\Gamma t}\, \bm{\tilde{s}}_2 &=& \tfrac{1}{2}\,(e^{s_+ t}\bm{s}_+ + 
                                       e^{s_+^* t}\bm{s}_+^*)
   = \mathrm{Re}\,[\,e^{s_+ t}\bm{s}_+\,]   \nonumber \\
&=& e^{-(\bar{R} + z_1/2)\,t}\,\mathrm{Re}\,
        [\,e^{i\varpi t}(\bm{\tilde{s}}_2 + i\,\bm{\tilde{s}}_3)\,]   \nonumber \\
&=& e^{-(\bar{R} + z_1/2)\,t}\,(\,\cos\varpi t\,\bm{\tilde{s}}_2 -
                              \sin\varpi t\,\bm{\tilde{s}}_3\,)\,.
\label{Prop s2}
     \end{eqnarray}
Similarly,
     \begin{eqnarray}
e^{-\Gamma t}\, \bm{\tilde{s}}_3 &=& -\tfrac{i}{2}\,(e^{s_+ t}\bm{s}_+ - 
                                       e^{s_+^* t}\bm{s}_+^*)
   = \mathrm{Im}\,[\,e^{s_+ t}\bm{s}_+\,]   \nonumber \\
&=& e^{-(\bar{R} + z_1/2)\,t}\,\mathrm{Im}\,
       [\,e^{i\varpi t}(\bm{\tilde{s}}_2 + i\,\bm{\tilde{s}}_3)\,]   \nonumber \\
&=& e^{-(\bar{R} + z_1/2)\,t}\,(\,\sin\varpi t\,\bm{\tilde{s}}_2 +
                              \cos\varpi t\,\bm{\tilde{s}}_3\,)\,.
            \nonumber \\
& &
\label{Prop s3}
     \end{eqnarray}
These relations, together with 
$e^{-\Gamma\,t}\bm{\tilde{s}}_1 = e^{s_1}\bm{\tilde{s}}_1$, yield the  propagator $e^{-\tilde{\Gamma}\,t}$ for the evolution of states
$\tilde{\mathcal{M}} = \sum_i \tilde{\mathcal{M}}_i \bm{\tilde{s}}_i$ expressed in the $\{\bm{\tilde{s}}_i\}$ basis, as given in \Eq{RotPictSoln}.

As noted in \Eq{BasisTrans}, matrix $P$ generated from the $\{\bm{\tilde{s}}_i\}$ entered as column vectors transforms from the
$\{\bm{\tilde{s}}_i\}$ basis to the standard basis, with 
$P^{-1} = \mathrm{adj}\,P/\det P$ giving the desired $\tilde{\mathcal{M}}$ starting with $\mathcal{M}$ in the standard basis. One easily shows that
$\det P = \bm{\tilde{s}}_1\cdot(\bm{\tilde{s}}_2\times\bm{\tilde{s}}_3)$, and 
row $i$, column $l$ of $\mathrm{adj}\,P$ is $(\bm{\tilde{s}}_j\times\bm{\tilde{s}}_k)_l$ for cyclic permutation of $i=1,j=2,\mathrm{and}\ k=3$ to obtain
     \begin{equation}
P^{-1} = \frac{1}
{\bm{\tilde{s}}_1\cdot(\bm{\tilde{s}}_2\times\bm{\tilde{s}}_3)}\,
 \left[\begin{array}{ccc}
     \cdots & (\bm{\tilde{s}}_2\times\bm{\tilde{s}}_3) & \cdots     \\
     \cdots & (\bm{\tilde{s}}_3\times\bm{\tilde{s}}_1) & \cdots     \\
     \cdots & (\bm{\tilde{s}}_1\times\bm{\tilde{s}}_2) & \cdots 
         \end{array}
  \right]
\label{Pinv}
     \end{equation}

The eigenvectors needed to construct the real basis are obtained in the usual fashion as solutions to 
$(s\Id + \Gamma)\bm{s} = 0$ for each eigenvalue $s_i$.  The  solution for the three components of each eigenvector is overdetermined, by construction, so any one of the three equations is a linear combination of the other two and is redundant.  We are free to assign any (nonzero) value to one of the components, leaving two equations and two unknowns.  There are three different but equivalent forms for the eigenvector solution depending on which two equations are chosen. Setting the third component equal to one for simplicity gives an expression for the other two components involving a common denominator.  Scaling the result by this factor gives the following result for eigenvector $\bm{s}_i$, with the left arrow signifying that the columns of the matrix map to $\bm{s}_i$:
\begin{widetext}
     \begin{equation}
\bm{s}_i \leftarrow \left[ \begin{array}{ccc}
   \omega_1^2 + (s_i+R_2)(s_i+R_3) & \ \ \omega_1\omega_2-\omega_3(s_i+R_3) &
      \ \ \omega_1\omega_3+\omega_2(s_i+R_2)  \\
  \omega_1\omega_2+\omega_3(s_i+R_3) & \ \  \omega_2^2 + (s_i+R_1)(s_i+R_3) &
     \ \ \omega_2\omega_3-\omega_1(s_i+R_1)  \\
  \omega_1\omega_3-\omega_2(s_i+R_2) &  \ \ \omega_2\omega_3+\omega_1(s_i+R_1) &
      \ \ \omega_3^2 + (s_i+R_1)(s_i+R_2) \\
                     \end{array}
             \right].
\label{EigVecMatrix}
     \end{equation}
\end{widetext}

The different columns give equivalent results, as discussed in section \ref{Propagator Dynamics}.  In the absence of relaxation, the real root of \Eq{CharPoly} is $s_1 = 0$ with eigenvector 
$\bm{s}_1 = (\omega_1,\omega_2,\omega_3)$, which is the rotation axis for the resulting time evolution.  In the case 
$\bm{\omega}_e = 0$, in which $\Gamma$ is already diagonal, the coordinates reduce to the standard coordinate system as required.

One might recognize the righthand side of \Eq{EigVecMatrix} as 
$\mathrm{adj\,}A(s_i)$ from Eqs.~(\ref{adjA_Poly}--\ref{adjA_coeff}), with
$\mathrm{adj}\,A(s_i) = \mathrm{adj}\,A_\p(z_i)$, since
$s_i = z_i - \bar R$ and $R_i - \bar R = R_{i\p}$.  We thus have the real basis vectors $\bm{\tilde{s}}_{2,3} \equiv \bm{\tilde{z}}_{2,3}$ equal to the respective real, imaginary parts of 
$\bm{z}_+ = \mathrm{adj}\,A_\p(z_+)$ according to \Eq{Real s_2,3}, with $z_+ = -z_1/2 + i\,\varpi$.  Then, using \Eq{adjAp_Poly} for $\mathrm{adj}\,A_\p(z_i)$ in polynomial form and eliminating common scale factors, the real basis vectors defining the oblique coordinate system can be written concisely as
     \begin{eqnarray}
\bm{\tilde{s}}_1 = \bm{\tilde{z}}_1 &\leftarrow& 
           A_{0\p} + A_{1\p}\,z_1 + \Id\,z_1^2  \nonumber \\
\bm{\tilde{s}}_2 = \bm{\tilde{z}}_2 &\leftarrow& 
           A_{0\p} - A_{1\p}\,\frac{z_1}{2} + 
           \Id\,\Big[\Big(\frac{z_1}{2}\Big)^2 - \varpi^2\,\Big]  \nonumber \\
\bm{\tilde{s}}_3 = \bm{\tilde{z}}_3 &\leftarrow&  A_{1\p} - \Id\,z_1
\label{RealBasis_zi}
     \end{eqnarray}
The result for $\bm{\tilde{z}}_1$ can be obtained directly from \Eq{EigVecMatrix} with the substitutions $s_i \rightarrow z_i$ and $R_i \rightarrow R_{ip}$ for the corresponding parameters associated with $\Gamma_\p$.  One can readily deduce the coefficient matrices $A_{0\p}$ and $A_{1\p}$ from \Eq{EigVecMatrix} and the expression for $\bm{\tilde{s}}_1$ in \Eq{RealBasis_zi} without recourse to the definitions for each element given in \Eq{adjAp_Poly}.  The matrices are also given as simple functions of $\Gamma_\p$ in \Eq{adjAp_coeff_Gamma}. For convenient reference, each coefficient matrix is written below.
     \begin{eqnarray}
A_{0\p} &=& \left[ \begin{array}{ccc}
   \omega_1^2 + R_{2\p}R_{3\p} & \  \omega_1\omega_2-\omega_3 R_{3\p} &
      \  \omega_1\omega_3+\omega_2 R_{2p}  \\
  \omega_1\omega_2+\omega_3 R_{3\p} & \  \omega_2^2 + R_{1\p}R_{3\p} &
     \ \omega_2\omega_3-\omega_1 R_{1\p}  \\
  \omega_1\omega_3-\omega_2 R_{2\p} &  \ \omega_2\omega_3+\omega_1 R_{1\p} &
      \ \omega_3^2 + R_{1\p}R_{2\p} \\
                     \end{array}
             \right]  \nonumber \\
& &  \nonumber \\
A_{1\p} &=& -\Gamma_\p = \left[ \begin{array}{ccc}
   -R_{1\p} & \ -\omega_3 & \ \omega_2  \\
  \omega_3 & \ -R_{2\p} & \ -\omega_1  \\
  -\omega_2 &  \ \omega_1 & \ -R_{3\p} \\
                     \end{array}
             \right],
\label{A0,A1}
     \end{eqnarray}
with $-R_{1\p} = R_{2\p} + R_{3\p}$ and cyclic permutations, since
$\sum_i R_{i\p} = 0$ by construction in the original matrix partitioning.

\subsection{Measures of obliquity}

Bloch equation dynamics are simple in the oblique coordinates of the model, consisting of independent rotation relaxation elements.  This section provides examples that quantify the degree to which the plane of rotation is oblique to the axis $\bm{\tilde z}_1$ representing simple exponential decay. In what follows, the first column of $\mathrm{adj}\,A_\p$ is arbitrarily chosen to calculate the coordinate basis $\{\bm{\tilde z}_i\}$ in the case $R_1 = R_2$.  Similar results are obtained using any of the other columns.

\subsubsection{\bf{Off resonance, $\bm{\omega}_e = (0, \omega_2, \omega_3)$}}

Off resonance, in contrast to the on-resonance example of section \ref{Onres}, $\bm{\tilde z}_1$ is neither aligned with $\bm{\omega}_e$, nor is it orthogonal to the $(\bm{\tilde z}_2,\bm{\tilde z}_3)$-plane.  Calculating the 
$\bm{\tilde z}_i$ as above provides the normal to the plane, 
$\bm{\tilde n}_{23} = \bm{\tilde z}_2 \times \bm{\tilde z}_3$.  We then have
     \begin{equation}
\bm{\tilde z}_1 = \left( \begin{array}{c}
     (z_1 + R_\delta)(z_1 - 2 R_\delta) \\ 
     \omega_3(z_1 - 2R_\delta) \\
     -\omega_2 (z_1 + R_\delta)
                         \end{array} \right)  
\label{z1 OffRes w1=0}
     \end{equation}
and 
     \begin{equation}
\bm{\tilde n}_{23} = \left( \begin{array}{c}
     3\omega_2\omega_3 R_\delta \\ 
     -\omega_2\,(a - z_1 R_\delta + z_1^2 + R_\delta^2\,) \\
     -\omega_3\,(a + 2 z_1 R_\delta + z_1^2 + 4 R_\delta^2\,)
                         \end{array} \right),  
\label{n OffRes w1=0}
     \end{equation}
which bears little resemblance to $\bm{\tilde z}_1$.  Yet, scaling $\bm{\tilde z}_1$ by $f_s = -(\bm{\tilde n}_{23})_1 / (\bm{\tilde z}_1)_1$ from the first components gives, for components two and three, 
$f_s \bm{\tilde z}_1 - \bm{\tilde n}_{23} \propto q(z_1)$, the characteristic polynomial for $-\Gamma_\p$, which is zero when evaluated at its root $z_1$.  Thus, within a scale factor or, equivalently, when both both vectors are normalized, we can write simply
     \begin{equation}
\bm{\tilde n}_{23} = \left( \begin{array}{c}
     -(\bm{\tilde z}_1)_1 \\ (\bm{\tilde z}_1)_2 \\ (\bm{\tilde z}_1)_3
                         \end{array} \right).  
\label{Scaled n OffRes w1=0}
     \end{equation}

\subsubsection{\bf{Off resonance, $\bm{\omega}_e = (\omega_1,0, \omega_3)$}}

Similarly, for $\omega_2 = 0$,
     \begin{equation}
\bm{\tilde z}_1 = \left( \begin{array}{c}
     \omega_1^2 + (z_1 + R_\delta)(z_1 - 2 R_\delta) \\ 
     \omega_3(z_1 - 2R_\delta) \\
     \omega_1 \omega_3
                         \end{array} \right)  
\label{z1 OffRes w2=0}
     \end{equation}
and 
     \begin{equation}
\bm{\tilde n}_{23} = -\left( \begin{array}{c}
     \omega_1\omega_3 \\ 
     \omega_1\,(z_1 + R_\delta) \\
     \tfrac{1}{4}(z_1 + 4R_\delta)^2 + \varpi^2 - \omega_1^2
                         \end{array} \right),  
\label{n OffRes w2=0}
     \end{equation}
Scaling $\bm{\tilde z}_1$ by $f_s = -(\bm{\tilde n}_{23})_2 / (\bm{\tilde z}_1)_2$ gives $f_s \bm{\tilde z}_1 - \bm{\tilde n}_{23} \propto q(z_1)$ for components one and three, so that
     \begin{equation}
\bm{\tilde n}_{23} = \left( \begin{array}{c}
     (\bm{\tilde z}_1)_1 \\ -(\bm{\tilde z}_1)_2 \\ (\bm{\tilde z}_1)_3
                         \end{array} \right) 
\label{Scaled n OffRes w2=0}
     \end{equation}
within a scale factor.

\subsubsection{$\bm{\omega_1 = \omega_2 = \omega_3 \equiv \omega}$}

In this case,
     \begin{equation}
\bm{\tilde z}_1 = \left( \begin{array}{c}
     \omega^2 + (z_1 + R_\delta)(z_1 - 2 R_\delta) \\ 
     \omega (\omega + z_1 - 2R_\delta) \\
     -\omega (\omega + z_1 + R_\delta)
                         \end{array} \right)  
\label{z1 w1=w2=w3}
     \end{equation}
and 
     \begin{equation}
\bm{\tilde n}_{23} = -\left( \begin{array}{c}
     \omega (2\omega - 3 R_\delta) \\ 
     \tfrac{1}{4}(z_1 - 2 R_\delta)^2 + \omega(z_1+R_\delta) +
      \varpi^2 - \omega^2 \\
     \tfrac{1}{4}(z_1 + 4 R_\delta)^2 - \omega(z_1+R_\delta) +
      \varpi^2 - \omega^2
                         \end{array} \right).  
\label{n w1=w2=w3 }
     \end{equation}
Scaling $\bm{\tilde z}_1$ by $f_s = (\bm{\tilde n}_{23})_1 / (\bm{\tilde z}_1)_2$ gives both $f_s (\bm{\tilde z}_1)_1 - (\bm{\tilde n}_{23})_2$  and
$f_s (\bm{\tilde z}_1)_3 - (\bm{\tilde n}_{12})_3$ proportional to $q(z_1)$, so that the vectors can be scaled to satisfy
     \begin{equation}
\bm{\tilde n}_{23} = \left( \begin{array}{c}
     (\bm{\tilde z}_1)_2 \\ (\bm{\tilde z}_1)_1 \\ (\bm{\tilde z}_1)_3
                         \end{array} \right).  
\label{Scaled n w1=w2=w3}
     \end{equation}

\section{An Alternative Method for Calculating \\ an Eigenvector }
\label{EigVecCalc}

Equation (\ref{EigVecMatrix}) is simply $\mathrm{adj}\, (s_i\Id + \Gamma)$ from Eqs.~(\ref{adjA}--\ref{adjA_coeff}).
One therefore happens upon the modest result, apparently unrecognized, that an eigenvector $\bm{\upsilon}$ corresponding to a distinct eigenvalue $\upsilon$ of operator $\Upsilon$ can be obtained as
     \begin{equation}
\bm{\upsilon} \in \mathrm{adj}\,(\upsilon\Id - \Upsilon),
\label{EigVecTh}
     \end{equation}
seen as follows.  Recall, the characteristic polynomial
$p(s) = \det(s\Id - \Upsilon)$ equals zero for eigenvalue $s = \upsilon$, and
$(s\Id - \Upsilon)^{-1} = \mathrm{adj}\,(s\Id - \Upsilon)/p(s)$ 
from \Eq{Ainv(s)}.  Then
     \begin{eqnarray}
p(s) &=& (s\Id - \Upsilon)\, \mathrm{adj}\,(s\Id - \Upsilon)  \nonumber \\
0 &=&  (\upsilon\Id - \Upsilon)\, 
                     \mathrm{adj}\,(\upsilon\Id - \Upsilon) \nonumber \\
\therefore \ \ \Upsilon\,\mathrm{adj}\,(\upsilon\Id - \Upsilon) &=&
\upsilon\,\mathrm{adj}\,(\upsilon\Id - \Upsilon)
\label{EigVecThPf}
     \end{eqnarray}
Only a single column of the adjugate matrix is required, so the method is fairly efficient.  However, the trivial zero eigenvector solution can be one of the columns, requiring further completion of the adjugate to obtain the desired eigenvector.  

For the case of degenerate eigenvalues, the method is incomplete. When the nullity (dimension of the null space) of 
$(\upsilon\Id - \Upsilon)$ equals the order of the degeneracy, $k$ (i.e, the rank equals the dimension of the operator, $n$, minus $k$), there are $k$ distinct eigenvectors, but the method fails, returning only the zero eigenvector.  
If there is not a complete set of eigenvectors (the degenerate eigenvalue is defective in that the nullity is less than $k$), and the rank is greater than 
$n-k$), the method appears to return the eigenvectors that exist, but one rarely needs these, since the matrix $\Upsilon$ is not diagonalizable in this case. 

\section{Limiting Cases}
The solutions are evaluated and confirmed for $R_1 = R_2$ and a representative set of limiting cases that can be readily solved  by other methods. 

\subsection{\textbf{Three distinct roots}}
Three examples are presented representing the separate cases $a=0$ and $b=0$.
\squishlist
\item[(i)] $b = 0$, $a \neq 0$
\squishend
According to the defining relations for $a$ and $b$ in \Eq{CardanCoeff3}, the condition $b=0$ implies
$\omega_{12}^2 = 2R_\delta^2(1+\frac{1}{3}\lambda_3 )$, using \Eq{omega_mag} for $\omega_e^2$ and \Eq{Param_w} for $\omega_3$.  Then
    \begin{equation}
a = \left\{ \begin{array}{ll}
   R_\delta^2(\lambda_3 - 1) & R_\delta \neq 0 \\
   \omega_e^2 & R_\delta = 0
            \end{array}
    \right .
\label{a(b=0)}      
     \end{equation}
The roots of \Eq{SCForm} are easily obtained, giving 
     \begin{eqnarray}
z_1 &=&  0 \qquad\qquad \varpi = \sqrt{a}. 
\label{params:b=0}
     \end{eqnarray}
There are two cases, depending on the sign of $a$.
\squishlist
\item[]
   \squishlist
      \item[] (1) $a > 0$
   \squishend
\squishend
Then \Eq{MatExp1} gives
     \begin{eqnarray}
e^{-\Gamma_\p\,t} &=& \
        \Id -\frac{\Gamma_\p}{\varpi}\,\sin\varpi t + 
      \left(\frac{\Gamma_\p}{\varpi}\right)^2\,(1-\cos\varpi t).
\label{Soln:a>0,b=0}
     \end{eqnarray}
There is no exponential decay contribution due to this term, with the overall factor $e^{-\bar R t}$ in the final expression for $e^{-\Gamma t}$ providing a single
system decay rate $\bar R$.
\vskip 6pt
\noindent \textit{Example (1)}
   \squishlist
      \item[\phantom{(***)}] Choose $R_\delta = 0$ to obtain
      \item[\phantom{(***)}]  $b=0$, \ \ $a = \omega_e^2$, \ \ 
                              $\varpi = \omega_e$
   \squishend
In this case, \Eq{Soln:a>0,b=0} represents a rotation about the field $\bm{\omega}_e$.

The propagator $U_R$ for a rotation about $\bm{\omega}_e$ is readily obtained by transforming to a coordinate system with new $z$-axis aligned with $\bm{\omega}_e$,  rotating by angle $-\omega_e t$ about this axis, then transforming back to the original coordinates.  Specifying the orientation of $\bm{\omega}_e$ in terms of polar angle $\theta$ and azimuthal angle $\phi$ relative to the $z$- and $x$-axes, respectively, one has 
$U_R = U_z(-\phi)U_y(-\theta)U_z(-\omega_e\,t)U_y(\theta) U_z(\phi)$ in terms of the elementary operators $U_y$ and $U_z$ for rotations about the $y$- and $z$- axes, respectively.  Then $U_R$ provides a verification of the \Eq{Soln:a>0,b=0} result upon substituting $\cos\phi = \omega_1/\omega_{12}$, $\sin\phi = \omega_2/\omega_{12}$,
$\cos\theta = \omega_3/\omega_e$, $\sin\theta = \omega_{12}/\omega_e$.

\squishlist
\item[]
   \squishlist
      \item[] (2) $a < 0$
   \squishend
\squishend
for $\lambda_3 < 1$ gives $\varpi \rightarrow i\,\mu = i\,\sqrt{|a|}$ and
     \begin{equation}
e^{-\Gamma_\p\,t} = \Id -\frac{\Gamma_\p}{\mu}\,\sinh\mu t + 
    2\left(\frac{\Gamma_\p}{\mu}\right)^2\,(\cosh\mu t - 1)
\label{Soln:a<0,b=0}
     \end{equation}
\vskip 6pt
\noindent \textit{Example (2)}
   \squishlist
      \item[\phantom{(***)}] Choose $\omega_1^2 = 2R_\delta^2$,\  
            $\omega_2 = 0$, $\lambda_3 = 0$ to obtain
      \item[\phantom{(***)}]  $b=0$, \ \ $a = -R_\delta^2$, \ \ 
                              $\mu = R_\delta$
   \squishend
Equation (\ref{Soln:a<0,b=0}) then gives
     \begin{eqnarray}
\lefteqn{e^{-\Gamma_\p\, t} = } \nonumber \\
& & \left(
   \begin{array}{ccc}
 e^{-R_\delta t} & 0 & 0 \\
 0 & 2 - e^{R_\delta t} & \sqrt{2}\,(1 - e^{R_\delta t})  \\
 0 & -\sqrt{2}\,(1 - e^{R_\delta t}) &  2\, e^{R_\delta t} - 1
      \end{array}
   \right). 
\label{MatExpEx1c}
     \end{eqnarray}

For an independent calculation, the matrix $-\Gamma_\p$ can be diagonalized, with eigenvalues given by the $z_i$ and associated real-valued eigenvectors. The simple exponential of the diagonalized matrix is then transformed back to the original basis in the standard fashion using the matrix of eigenvectors and its inverse to obtain $e^{-\Gamma_\p\, t}$ as given above.   

\squishlist
\item[(ii)] $a = 0$. $b \neq 0$
\squishend
The condition $a=0$ implies
$\omega_e^2 = 3R_\delta^2$, leading to
    \begin{equation}
b = R_\delta^3(1 - \lambda_3 )
\label{b(a=0)}      
     \end{equation}
and root $z_1 = -\mathrm{sgn}(b)|b|^{1/3}$ from \Eq{Root_z1:4}.  
For $\mathrm{sgn}(b) = \pm 1$ and the definition $\tilde \lambda_3 = |1-\lambda_3 |^{1/3}$, we have accordingly
     \begin{eqnarray}
z_1 &=& \mp\, \tilde\lambda_3 R_\delta \qquad\qquad 
        \varpi = \frac{\sqrt{3}}{\ 2}\,\tilde\lambda_3 R_\delta 
\label{params:a=0}
     \end{eqnarray}
Although the form of \Eq{MatExp1} does not simplify in this case as appreciably as for $b = 0$, both the root $z_1$, which determines the decay rate, and the oscillatory frequency $\varpi$ are simple multiples of $R_\delta$.  
\vskip 6pt
\noindent \textit{Example (3)}
   \squishlist
      \item[\phantom{(***)}] Choose 
           $\omega_e^2 \rightarrow\omega_1^2 = 3R_\delta^2$, 
            \qquad  $\omega_2 = 0 =\omega_3$
   \squishend
Most off-diagonal elements of $\Gamma_\p$ are equal to zero in this case, and $\tilde\lambda_3 = 1$ for the \Eq{params:a=0} input parameters to 
\Eq{MatExp1}.
Defining $\kappa = (\sqrt{3}/2)R_\delta$ and combining the sums of trigonometric functions that appear on the diagonal gives the succinct form
     \begin{eqnarray}
\lefteqn{e^{-\Gamma_\p\, t} = } \nonumber \\
& & e^{\frac{1}{2}R_\delta t}\,
\left(
   \begin{array}{ccc}
 e^{-\frac{3}{2} R_\delta t} & 0 & 0 \\
 0 & -2 \sin \left(\kappa t - \frac{\pi }{6}\right) & 
    -2 \sin (\kappa t ) \\
 0 & 2 \sin (\kappa t) & 2 \sin \left(\kappa t +\frac{\pi}{6}\right) 
   \end{array}
\right) \nonumber \\
 & &
\label{MatExpEx1a}
     \end{eqnarray}

Again, the matrix $-\Gamma_\p$ is diagonalizable, providing a simple result for the matrix exponential in the eigenbasis and a straightforward means for calculating 
$e^{-\Gamma_p t}$ as obtained above.  The associated eigenvectors are complex-valued in this case, making the algebra slightly more tedious.  Alternatively, one can readily verify that $d/dt\,e^{-\Gamma_\p\, t} = -\Gamma_\p\, e^{-\Gamma_\p\, t}$.  

\subsection{\textbf{Two equal roots}}

Degenerate roots require $\gamma = 1$.  
For a given $\omega_3^2 = \lambda_3 R_\delta^2/3$, with $0\le\lambda_3 \le1$, there are two values $\omega_{12}^2$ that satisfy $\gamma = 1$.  Consider $\lambda_3 = 0$, in which case Eqs.~(\ref{Param_w}) and (\ref{etaSolns}) give
     \begin{eqnarray}
(\vartheta_1, \vartheta_2) &=& (\,-\pi/6\,,\, \pi/2\,)  \nonumber \\
(\eta_1, \eta_2) &=& (\,-9/4\,,\, 9/2\,)  \nonumber \\
(\omega_{12,1}^2\,,\,\omega_{12,2}^2) &=& (\,0\,,\,9/4 R_\delta^2\,)
\label{2 w_12 params}
     \end{eqnarray}
\squishlist
\item[(i)] $\omega_{12} = 0$ 
\squishend
This is the case of pure relaxation, with $\Gamma_\p$ reduced to the diagonal elements $\{R_\delta, R_\delta, -2R_\delta\}$.  We have $a = -3 R_\delta^2$,
$b = -2 R_\delta^3 < 0$, and
     \begin{eqnarray}
z_1 &=& 2 R_\delta \qquad\qquad 
        \varpi = 0 
\label{RealParamsEx2a} 
     \end{eqnarray}
from \Eq{Root_z1:2}.  Thus, \Eq{MatExp2} gives the expected result
    \begin{equation}
e^{-\Gamma_P\,t} = 
\left( \begin{array}{ccc}
  e^{-R_\delta t} & 0 & 0 \\
  0 & e^{-R_\delta t} & 0 \\
  0 & 0 & e^{2 R_\delta t} 
        \end{array}
\right).
\label{MatExpEx2a}      
     \end{equation}

\squishlist
\item[(ii)] $\omega_{12}^2 = \frac{9}{4}R_\delta^2 \rightarrow \omega_1^2$
\squishend
Then $a = -3 R_\delta^2/4 < 0$, $b = R_\delta^3/4 > 0$, and
     \begin{eqnarray}
z_1 &=& -R_\delta \qquad\qquad 
  \varpi = 0 
\label{RealParamsEx2b} 
     \end{eqnarray}
resulting in
     \begin{eqnarray}
\lefteqn{e^{-\Gamma_\p\, t} = } \nonumber \\
& & e^{\frac{1}{2}R_\delta t}\,
\left(
   \begin{array}{ccc}
 e^{-\frac{3}{2} R_\delta t} & 0 & 0 \\
 0 & 1-\omega_1 t & -\omega_1 t  \\
 0 & \omega_1 t & 1+\omega_1 t  
   \end{array}
\right). 
\label{MatExpEx2b}
     \end{eqnarray}
Verifying that $d/dt\,e^{-\Gamma_\p\, t} = -\Gamma_\p\, e^{-\Gamma_\p\, t}$ is fairly straightforward and represents the simplest test of the solution, since
$\Gamma_\p$ is not diagonalizable.

\subsection{\textbf{Three equal roots}}

There is a three-fold degenerate root $z_i=0$ in the case $a=0=b$, since
$q(z) \rightarrow z^3$. This requires $\omega_e^2 = 3R_\delta^2$ from \Eq{CardanCoeff3}, which then forces 
$\omega_3^2 = R_\delta^2/3$ in the expression for $b$.  As noted previously, the Cayley-Hamilton theorem is simple to apply directly in this case, since $q(\Gamma_\p) = \Gamma_\p^3 = 0$.  The series expansion of $e^{-\Gamma_\p\,t}$ is therefore truncated, giving the \Eq{MatExp3} result.

\subsection{\textbf{On resonance }}

When $\omega_3 = 0$, $b$ can be written in the form
$R_\delta(a + R_\delta^2)$ from \Eq{CardanCoeff3}, with
$a \rightarrow \omega_{12}^2 - 3 R_\delta^2$.  The characteristic polynomial then becomes
$z^3 + R_\delta^3 + a(z+R_\delta)$, so that, by inspection, 
     \begin{eqnarray}
 z_1 &=& -R_\delta \qquad\quad 
        \varpi = \sqrt{\omega_{12}^2 - (\tfrac{3}{2}R_\delta)^2}
\label{params:OnRes}
     \end{eqnarray}
The solution for $e^{-\Gamma_p \,t}$ using \Eq{MatExp1} with the above parameters yields the solution for $e^{-\Gamma\,t}$ obtained originally in \cite{Torrey} for the case $\varpi \neq 0$. As discussed above, if 
$\omega_{12} = 3R_\delta/2$, there is a two-fold degeneracy in the roots, giving the solution in \Eq{MatExpEx2b} for $e^{-\Gamma_p \,t}$.

For $\omega_{12} < 3R_\delta/2$, the sinusoidal terms become the corresonding hyperbolic functions, as noted earlier, with
$\cos\varpi\,t \rightarrow \cosh\mu\,t$ and
$\sin\varpi\,t/\varpi \rightarrow \sinh\mu\,t/\mu$, where now
$\mu = \sqrt{(\tfrac{3}{2}R_\delta)^2 - \omega_{12}^2}$.

\begin{widetext}
\par\vfill\eject
\end{widetext}
\bibliography{/home/tskinner/tex/texmf/BibTeX/bib-database-TES}
\bibliographystyle{apsrev4-1}

\par\vfill\eject

\par\vfill\eject

\par\vfill\eject

\par\vfill\eject

\end{document}